\documentclass[a4paper,11pt]{article}
\usepackage{epsfig}
\usepackage{axodraw}
\usepackage{graphicx}
\usepackage{amsmath}

\newcommand{\Slash}[1]{\ooalign{\hfil/\hfil\crcr$#1$}}

\begin{document}

\title{
CP Asymmetry, Branching ratios and Isospin 
breaking effects in $B \to \rho\gamma $ and 
$B \to \omega\gamma $ decays
with the pQCD approach
}

\author{
Cai-Dian L\"u$^{a,b}$\footnote{Electronic address: lucd@mail.ihep.ac.cn},
M. Matsumori$^{c}$\footnote{Electronic address: mika@eken.phys.nagoya-u.ac.jp},
A.I. Sanda$^{c}$\footnote{Electronic address: sanda@eken.phys.nagoya-u.ac.jp},
and Mao-Zhi Yang$^{a,b}$\footnote{Electronic address: yangmz@mail.ihep.ac.cn}
\\
{\small $^a$CCAST (World Laboratory), P.O. Box 8730, Beijing 100080,
     China;}\\
{\small $^b$Institute of High Energy Physics,
 P.O. Box 918(4), Beijing 100049, China;}\\
{\small $^c$Department of Physics, Nagoya University, Chikusa-Ku Furo-cho
    Nagoya 464-8602 Japan }
}

\date{}
\maketitle
\begin{abstract}
The radiative ${B\rightarrow \rho \gamma}$, ${B\to \omega \gamma}$ 
decay modes are caused by the flavor-changing-neutral-current
process, 
so they give us good insight towards probing
the standard model in order to
search for new physics.
In this paper, we compute the branching ratio, 
direct 
CP asymmetry, and isospin
breaking effects using the perturbative QCD approach
within the standard model.
\end{abstract}

\section{Introduction}
The standard model (SM) predicts large CP violation
in ${B}$ decays \cite{Carter:1980tk,Bigi:1981qs} and they have been
verified in
${B\to J/\psi K_s}$ \cite{Abe:2003yu,Raven:2003gs},
 ${B\to \pi\pi}$ \cite{Aubert:2004aq,Bevan:2004ht},
and ${B\to DK}$ \cite{Abe:2004gu}
decays. The quest of high energy physics has
always been to search for the most fundamental theory.
So our immediate goal is to search for deviation
from the predictions of the SM.
It is believed that the quantum effects in ${B}$ meson
decay amplitudes may contain effects of new physics.

The flavor-changing-neutral-current (FCNC) process
which causes ${b\to s\gamma}$ and ${b\to d\gamma}$
decays may contain new physics (NP) effects through 
penguin amplitudes.
As the SM effects represent the background
when we search for NP effects,
we shall compute these effects.
In doing so, we can understand the
sensitivity of each NP search. 

The first experimental evidence of this FCNC transition process 
in ${B}$ decay was
observed about a decade ago, where the inclusive process $b\to s
\gamma$ and exclusive process $B\to K^*\gamma$ were detected, and
their branching ratios were measured \cite{gamma}.
On the other hand,
the expected branching ratio for ${b\to d\gamma}$ 
is suppressed by ${O(10^{-2})}$
with respect to that for ${b\to s\gamma}$, 
because of the Cabbibo-Kobayashi-Maskawa (CKM)
quark-mixing matrix factor. 
The world average for ${b\to d}$ penguin decays
is given as follows
\cite{Group(HFAG):2005rb}:
\begin{equation*}
\begin{cases}
&Br(B^0\to \rho^0 \gamma )=(0.38\pm 0.18)\times 10^{-6}\\
&Br(B^0\to \omega \gamma )=(0.54^{+0.23}_{-0.21})\times 10^{-6}\\
&Br(B^+\to \rho^+\gamma )=(0.68^{+0.36}_{-0.31})\times 10^{-6}.
\end{cases}
\end{equation*}

Theoretically, $B\to\rho\gamma$ and $B \to\omega\gamma$ are widely
studied both within and beyond the SM
\cite{theory1,theory2}. The bound states are involved in
the exclusive process, so the perturbation theory can not be used in a
simple manner. 
It has been shown that, at least in the leading order,
all nonperturbative effects can be included
in the definition of the ${B}$ meson 
and the vector meson wave functions,
and the rest of the amplitude (the hard part
of the amplitude) can be computed in the perturbation theory.
This is called 
the perturbative QCD (pQCD) approach and it was proven several
years ago
\cite{li1,li2}.
In this paper, we compute 
the branching ratio, direct CP asymmetry,
and isospin breaking effects for
${B\to \rho\gamma}$, ${B\to \omega\gamma}$
decays
by using the 
pQCD
within the SM.

The remaining part of this paper is organized as follows. In
Sec.\ref{pQCD},
we briefly review the pQCD approach,
and in Sec.\ref{formula}, we present
 some basic formulas such as the effective Hamiltonian and
kinetic conventions. In
Sec.\ref{calculation}, the hard amplitudes  calculated in pQCD are
given. Section \ref{numerical} is devoted to numerical calculation and
discussion. Finally, a brief summary is given in Sec.\ref{conclusion}.

\section{Perturbative QCD Approach}
\label{pQCD}
\begin{figure}
\begin{center}
\includegraphics[width=6cm]{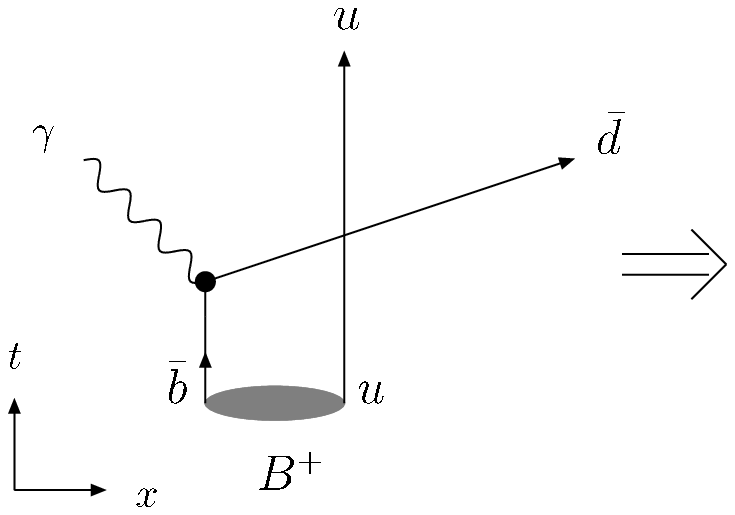}
\hspace{1mm}
\includegraphics[width=5.5cm]{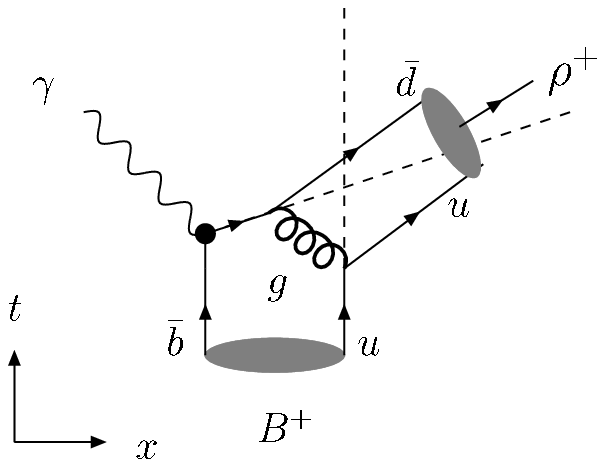}
 \caption{The heavy ${\bar{b}}$ quark decays into
the light ${\bar{d}}$ quark and a photon through the
electromagnetic operator, and the decay products
dash away 
back-to-back 
with momenta ${O(M_B/2)}$.
In order to form a ${\rho^+}$ meson with no hadron jets,
the spectator quark
must line up with ${\bar{d}}$. This can be accomplished
most efficiently by exchanging a hard gluon}
 \label{gluon}
\end{center}
\end{figure}

In order to explain the
pQCD approach,
we want to 
suppose that a
static ${B^+}$ meson decays
into ${\rho^+}$ and ${\gamma}$ through 
the ${O_{7\gamma}}$ operator as in Fig.\ref{gluon}.

\begin{figure}
\begin{center}
\includegraphics[width=7.7cm]{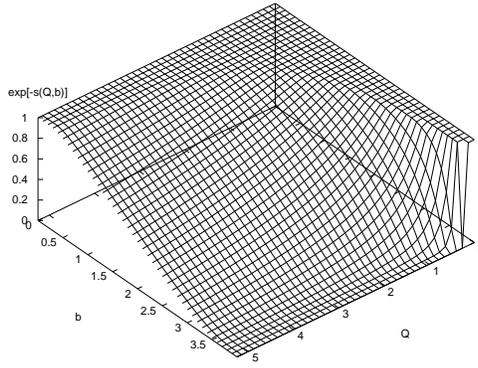}
\caption{The dependence of the Sudakov factor ${\exp[-s(Q,b)]}$
on ${Q}$ and ${b}$ where ${Q}$ is the ${b}$ quark momentum,
and ${b}$ is the interval between quarks
which form hadrons. It is clear that the large ${b}$ and ${Q}$
region is
suppressed.}
\label{Sudacov}
\end{center}
\end{figure}

\begin{figure}
 \begin{center}
\includegraphics[width=2.8cm]{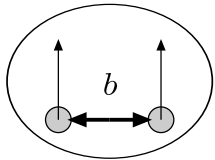}
 \end{center}
 \caption{${b}$ is the transverse interval between the ${\bar{b}}$
 and ${u}$ quark in the ${B}$ meson.}
 \label{interval}
\end{figure}

In the rest frame of the ${B^+}$ meson, the ${\bar{b}}$ quark is
almost at rest and the spectator ${u}$ quark moves around
the ${\bar{b}}$ quark
with ${O(\Lambda)=O(M_B-m_b)}$ momentum,
where ${M_B}$, and ${m_b}$ are ${B}$ meson,
and ${b}$ quark mass, respectively. 
Then the ${\bar{b}}$ quark decays into ${\bar{d}}$ and ${\gamma}$,
and these products dash away 
back-to-back 
with ${O(M_B/2)}$ momenta.
When a quark is rapidly accelerated like this,
infinitely many gluons are likely to be emitted
by bremsstrahlung.
There is a familiar 
phenomena in QED,
when an electrically charged particle
is accelerated, infinitely many photons are emitted.
But the gluon emission by bremsstrahlung QCD
must result in
many hadrons in the final state.
As the emitted gluon will hadronize,
the fact that no hadron except for ${\rho(\omega)}$
should be observed in ${B\to \rho(\omega)\gamma}$,
means that
the bremsstrahlung gluon emission
mentioned above can not occur.
Thus the branching ratio for an exclusive decay
${B\to \rho(\omega)\gamma}$ is proportional to the
probability that no bremsstrahlung gluon is emitted.
The amplitude for an exclusive decay contains 
the Sudakov factor and it is depicted in
Fig.\ref{Sudacov}.
As seen in Fig.\ref{Sudacov}, the Sudakov factor
is large for small ${b}$ and small ${Q}$,
where ${b}$ is the spacial distance between quark and antiquark
into ${B}$ meson, as shown in 
Fig.\ref{interval}, and ${Q}$ is the ${b}$ quark momentum
inside the ${B}$ meson.
Large ${b}$ implies that the quark and antiquark pair is separated
in space,
which in turn implies less color shielding.
Similar absence of the color shielding occurs 
when the ${b}$ quark carries the most of the momentum of the ${B}$
meson.
That is, as seen in Fig.\ref{Sudacov},
in order to form a ${\rho(\omega)}$ meson with no hadron jets,
the condition for color shielding is essential.
The condition needed for the color shielding is the
small separation in space between quark and antiquark 
within the meson,
and it indicates that the energy scale of the decay process
should be high. 
Actually,
the invariant-mass square of the
exchanged gluon depicted in Fig.\ref{gluon} is about
${O(\Lambda M_B)}$,
which can be considered to be in the short distance regime.
Thus we can see that the decay process
can be treated perturbatively.
The decay amplitude for the exclusive mode
like ${B\to \rho(\omega)\gamma}$ decay
can be factorized into the hard part
with a hard gluon exchange, which can be
treated perturbatively,
and the soft part 
of all nonperturbative strong interactions
is included in the meson wave functions.

Then the total decay amplitude can be
expressed as the convolution  like
\begin{eqnarray}
&&\hspace{-5mm}\int^1_0 dx_1 dx_2 \int^{1/\Lambda}_0 d^2b_1
 d^2b_2\hspace{1mm}C(t) \otimes \exp{[-S(x_1,x_2,b_1,b_2,t)]}\nonumber\\
&&\hspace{2.5cm}
\otimes
\Phi_{\rho,\hspace{1mm}\omega}(x_2,b_2)
\hspace{1mm}\otimes H(x_1,x_2,b_1,b_2,t)\otimes\Phi_B(x_1,b_1),
\end{eqnarray}
where ${\Phi_{\rho\hspace{1mm}\omega}(x_2,b_2)}$ and ${\Phi_B(x_1,b_1)}$ are
meson distribution amplitudes,
${\exp{[-S(x_1,x_2,b_1,b_2,t)]}}$ is the Sudakov factor,
which results from summing
up all the double logarithms of the 
soft divergences.
${H(x_1,x_2,b_1,b_2,t)}$
is the hard kernel including finite piece of quantum correction,
${b_1}$, ${b_2}$ are the conjugate variables
to transverse momenta, and 
${x_1}$, ${x_2}$ are the momentum fractions
of spectator quarks.

In the computation of the decay amplitudes
with the pQCD approach, we adopt the model functions
for the meson distribution amplitudes.
The meson amplitudes are  characterized
by the strong interaction.
The effective range
of the strong interaction which
can propagate,
is wide.
Then the meson distribution amplitudes
should be expressed as some averaged physical quantity.
Thus the meson amplitude does not depend on 
the decay process etc. 
For the ${B}$ meson wave function, 
we adopt a model \cite{pQCD}.
For
the ${\rho}$ and ${\omega}$ meson wave function,
we use ones
determined by the light-cone QCD sum rule \cite{ball}.
The detailed expressions for the meson functions are in Appendix
\ref{C}.

\section{Basic formulas}
\label{formula}
The flavor-changing ${b\rightarrow d \gamma}$
transition induced by an effective Hamiltonian
is given by \cite{buras}
\begin{eqnarray}
H_{\mbox{\scriptsize{eff}}}&=&\frac{G_F}{\sqrt 2}
\Big[
V_{ub}V_{ud}^*\left\{
C_1^{(u)}(\mu)O_1^{(u)}(\mu)+C_2^{(u)}(\mu)O_2^{(u)}(\mu)\right\}
\nonumber\\
&&\hspace{6mm}
+V_{cb}V_{cd}^*\left\{
C_1^{(c)}(\mu)O_1^{(c)}(\mu)+C_2^{(c)}(\mu)O_2^{(c)}(\mu)\right\}\\
&&\hspace{6mm}
-V_{tb}V_{td}^*\Big\{
\sum_{i=3\sim 6}C_{i}(\mu)O_{i}(\mu)+C_{7\gamma}(\mu)O_{7\gamma}(\mu)
+C_{8g}(\mu)O_{8g}(\mu)\Big\}
\Big]+(\mbox{h.c.}),\nonumber
\end{eqnarray}
where $C_i$'s are Wilson coefficients, and $O_i$'s are local operators
which are given by
\begin{eqnarray}
\label{operator}
O_1^{(q)}&=&(\bar{d}_i q_j)_{V-A}(\bar{q}_j b_i)_{V-A},
\hspace{2.6cm}
O_2^{(q)}=(\bar{d}_i q_i)_{V-A}(\bar{q}_j b_j)_{V-A},\nonumber\\
O_3^{(q)}&=&(\bar{d}_i b_i)_{V-A}\sum_q(\bar{q}_j q_j)_{V-A},
\hspace{2cm}
O_4^{(q)}=(\bar{d}_i b_j)_{V-A}\sum_q(\bar{q}_j q_i)_{V-A},
\nonumber\\
O_5^{(q)}&=&(\bar{d}_i b_i)_{V-A}\sum_q(\bar{q}_j q_j)_{V+A},
\hspace{2cm}
O_6^{(q)}=(\bar{d}_i b_j)_{V-A}\sum_q(\bar{q}_j q_i)_{V+A},\\
O_{7\gamma}&=&\frac{e}{8{\pi}^2}m_b\bar{d}_i\sigma^{\mu \nu}(
 1+\gamma_5)b_iF_{\mu \nu},
\hspace{16mm}
O_{8g}=\frac{g}{8{\pi}^2}m_b\bar{d}_i\sigma^{\mu \nu}(
 1+\gamma_5)T_{ij}^ab_jG_{\mu \nu}^a, \nonumber
\end{eqnarray}
and we neglect the terms which are proportional to
${d}$ quark mass in ${O_{7\gamma}}$ and ${O_{8g}}$.
Here ${(\bar{q}_iq_j)_{V\mp A}}$ means 
${\bar{q}_i\gamma^{\mu}(1\mp \gamma^5)q_j}$,
and ${i}$, ${j}$ are color indexes.
With the effective Hamiltonian given above, the decay amplitude of
$B\to \rho (\omega )\gamma$ can be expressed as
\begin{equation}
A=\langle F|H_{\mbox{\scriptsize{eff}}}|B\rangle=\frac{G_F}{\sqrt{2}}\sum_{i,q}V_{qb}^*V_{qd}
C_i(\mu)\langle F|O_i (\mu)|B\rangle ,
\end{equation}
where $F$ denotes the final state $\rho\gamma$ or $\omega\gamma$.
In addition, the amplitude can be decomposed into scalar (${M^S}$) and
pseudo-scalar (${M^P}$) components as
\begin{equation}
A=(\varepsilon_{V}^*\cdot\varepsilon_{\gamma}^*)M^S+
\frac{i}{P_V\cdot P_{\gamma}}\epsilon_{\mu\nu\rho\sigma}
\varepsilon_{\gamma}^{*\mu}\varepsilon_{V}^{*\nu}
P_{\gamma}^{\rho}P_V^{\sigma}M^P, \label{MSP}
\end{equation}
where $P_V$, and $P_{\gamma}$ are the momenta of $\rho$($\omega$)
meson, and photon, respectively. $\varepsilon_{\gamma}^*$ and
$\varepsilon_{V}^*$ are the relevant
polarization vectors. The matrix element $\langle F|O_i
(\mu)|B\rangle $ can be calculated in the pQCD approach.

For convenience, we work in light-cone coordinate. Then the
momentum is taken in the form 
\begin{eqnarray}
p=(p^+,
 p^-,\vec{p}_{T})=\left(\frac{p^0+p^3}{\sqrt{2}},\frac{p^0-p^3}{\sqrt{2}},(p^1,p^2)\right),
\end{eqnarray}
and the scalar product of two arbitrary
vectors $A$ and $B$ is $A\cdot B=A_{\mu}B^{\mu}= 
(A^+B^- + A^-B^+) - \vec A_{\bot}\cdot \vec B_{\bot}$. In the
$B$ meson rest frame, 
the momentum of $B$ meson is
\begin{eqnarray}
P_B=(P_B^+,P_B^-,\vec{P}_{B\perp})=\frac{M_B}{\sqrt 2}( 1,1,\vec 0_\bot),
\end{eqnarray}
and by choosing the coordinate frame where the $\rho$ or $\omega$
meson moves in the ``-" and photon in the ``+" direction, the
momenta of final state particles are
\begin{eqnarray}
P_{\gamma} &=&(P_{\gamma}^+,P_{\gamma}^-,\vec{P}_{\gamma\perp})= \frac{M_B}{\sqrt 2}( 1,0,\vec 0_\bot),\\
P_V &=&(P_V^+,P_V^-,\vec{P}_{V\perp}) =\frac{M_B}{\sqrt 2}( 0,1,\vec
 0_\bot).
\end{eqnarray}
The momenta of the spectator quarks in $B$ and
$\rho$ or $\omega$ mesons are
\begin{eqnarray}
k_1&=&(k_1^{+},k_1^{-},\vec k_{1T})=(\frac{M_B}{\sqrt 2}x_1,0,\vec{k_{1T}}),\\
k_2&=&(k_2^{+},k_2^{-},\vec k_{2T} )=(0,\frac{M_B}{\sqrt 2}x_2,\vec{k_{2T}}),
\end{eqnarray}
where $x_1$, and $x_2$ are momentum fractions which are defined by
$x_1=k_1^+/P_B^+$, and $x_2=k_2^-/P_V^-$, respectively.
  
\section{Formulas of the hard amplitude}
\label{calculation}

In this section we give the amplitudes caused by each operator
in Eq.(\ref{operator}).

\subsection{Contribution of $O_{7\gamma}$}

\begin{figure}
 \begin{center}
\includegraphics[width=5cm]{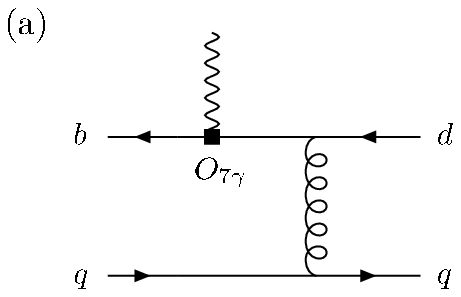}
\hspace{1cm}
\includegraphics[width=5cm]{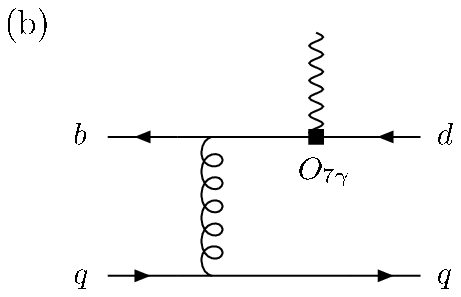}
  \end{center}
 \caption{Contribution from operator $O_{7\gamma}$ to
 $B\to\rho (\omega )\gamma $ decay. The photon
is emitted through the operator, and hard gluon exchange
is need to form a ${\rho (\omega)}$ meson.}
  \label{figo7}
\end{figure}

At first, we present the contribution of 
the electromagnetic operator $O_{7\gamma}$.
The diagrams are shown in Fig.\ref{figo7}.
In this case, the photon
is emitted through the operator, and hard gluon exchange
is needed to form a ${\rho (\omega)}$ meson.
 Contributions of the
$O_{7\gamma}$ operator to the amplitudes $M^S$ and $M^P$ defined in
Eq.(\ref{MSP}) are as follows:

\begin{eqnarray}
M_{7\gamma}^{S(a)}&=& -M_{7\gamma}^{P(a)}\nonumber\\
&=& 
-2F^{(0)}\xi_t\int dx_1dx_2\int
db_1db_2b_1b_2\alpha_s(t_7^a) 
\exp[-S_{B}(t_7^a)-S_V(t_7^a)]S_t(x_1)C_{7\gamma}(t_7^a)
\phi_{B}(x_1,b_1)\nonumber\\
&\times &
r_V\Big[\phi^v_V(x_2)+\phi^a_V(x_2)\Big]H_7^{(a)}(A_7b_2,B_7b_1,B_7b_2),
\nonumber\\
&&\hspace{3cm}
\left( t_7^a=\mbox{max}(A_7,B_7,1/b_1,1/b_2) \right),
\end{eqnarray}
\begin{eqnarray}
M_{7\gamma}^{S(b)}&=& -M_{7\gamma}^{P(b)}\nonumber\\
&=&
-2F^{(0)}\xi_t\int dx_1dx_2\int db_1db_2b_1b_2
\alpha_s(t_7^b)\exp[-S_{B}(t_7^b)-S_V(t_7^b)]S_t(x_2)C_{7\gamma}(t_7^b)
\phi_{B}(x_1,b_1)\nonumber\\
&\times &
 \Big[(1+x_2)\phi^T_V(x_2)
+(1-2x_2)r_V[\phi^a_V(x_2)+\phi^v_V(x_2)]\Big]
H_7^{(b)}(A_7b_1,C_7b_1,C_7b_2),\nonumber\\
&&
\hspace{3cm}\left(t_7^b
	       =\mbox{max}(A_7,C_7,1/b_1,1/b_2)\right),
\end{eqnarray}
\begin{eqnarray}
H_7^{(a)}\left(A_7b_2,B_7b_1,B_7b_2\right)
&\equiv &
K_0(A_7b_2)\Big[\theta(b_1-b_2)K_0\left(B_7b_1\right)I_0\left(B_7b_2\right)
\nonumber\\
&&\hspace{3cm}+\theta(b_2-b_1)K_0\left(B_7b_2\right)I_0\left(B_7b_1\right)\Big],
\end{eqnarray}
\begin{eqnarray}
H_7^{(b)}(A_7b_1,C_7b_1,C_7b_2) = H_7^{(a)}(A_7b_1,C_7b_1,C_7b_2),
\end{eqnarray}
\begin{eqnarray}
A_7^2= x_1x_2M_B^2,\hspace{5mm}B_7^2 = x_1{M_B}^2,\hspace{5mm}C_7^2=  x_2 M_B^2. 
\end{eqnarray}
Here ${K_0,~I_0}$ are modified Bessel functions which
are
extracted by the propagator integrations.
We define the common factor as
\begin{eqnarray}
F^{(0)}=\frac{G_F}{\sqrt{2}}\frac{e}{\pi}C_FM_B^5,
\end{eqnarray}
and the CKM matrix element as
${\xi_q=V_{qb}^*V_{qs}}$.
The exponentials
$\exp[-S_B(t)]$ and $\exp[-S_{V}(t)]$ are the Sudakov factors \cite{li1},
and
the explicit expressions of the exponents $S_B$, $S_{V}$ 
are shown in Appendix \ref{B}.
The quark structures for vector mesons
are ${\rho^+=|\bar{d}u\rangle}$, 
~${\rho^0=|\bar{u}u-\bar{d}d\rangle/\sqrt{2}}$,
~and ${\omega=|\bar{u}u+\bar{d}d\rangle/\sqrt{2}}$,
then the decay amplitudes for each decay modes
caused by ${O_{7\gamma}}$ operator
are given as follows:
\begin{eqnarray}
M(B^+\to
 \rho^+\gamma)_{7\gamma}^j&=&M_{7\gamma}^{j(a)}+M_{7\gamma}^{j(b)},\\
\nonumber\\
M(B^0\to
 \rho^0\gamma)_{7\gamma}^j&=&
-\frac{1}{\sqrt{2}}
\left[M_{7\gamma}^{j(a)}+M_{7\gamma}^{j(b)}\right],\\
M(B^0\to
 \omega\gamma)_{7\gamma}^j&=&\frac{1}{\sqrt{2}}
\left[M_{7\gamma}^{j(a)}+M_{7\gamma}^{j(b)}\right],
\end{eqnarray}
where ${j}$ expresses the decay amplitude components ${S}$ or ${P}$.

\subsection{Contribution of $O_{8g}$}

\begin{figure}
 \begin{center}
\vspace{5mm}
\includegraphics[width=4.5cm]{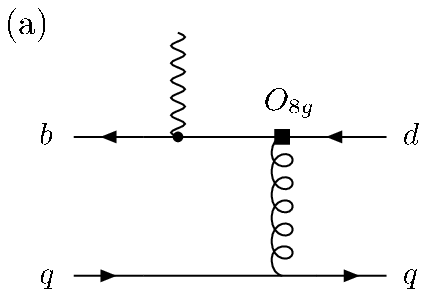}
\hspace{1cm}
\includegraphics[width=4.5cm]{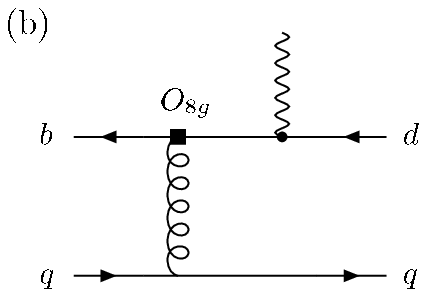}
\vspace{1cm}
\\
\includegraphics[width=4.5cm]{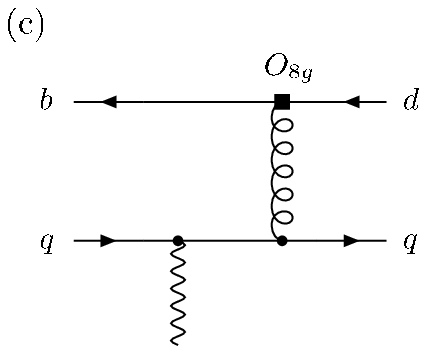}
\hspace{1cm}
\includegraphics[width=4.5cm]{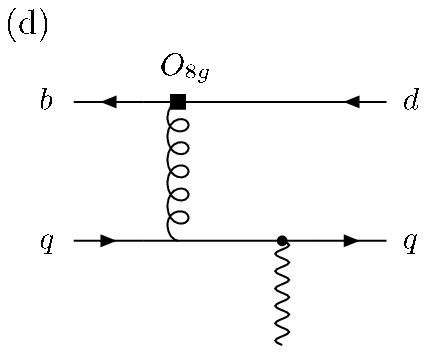}
  \end{center}
 \caption{Diagrams for the contribution of the
chromomagnetic operator $O_{8g}$. A hard gluon is 
emitted through the $O_{8g}$ operator and
glued to the spectator quark line.
Thus a  photon is emitted by the bremsstrahlung of the external quark lines.}
 \label{figo8g}
\end{figure}

The diagrams for the contribution of the chromomagnetic penguin operator
$O_{8g}$ are shown in Fig.\ref{figo8g}. Contributions of each
diagram are given in the following.
In this case, a hard gluon is emitted through the ${O_{8g}}$
operator and glued to the spectator quark line,
and a photon is emitted
by the bremsstrahlung from the external quark lines.
Each decay amplitude caused by
${O_{8g}}$ operator is expressed as follows:

\begin{eqnarray}
M_8^{S(a)}(Q_b)&=&-M_8^{P(a)}(Q_b)\nonumber\\
&=&
-F^{(0)}\xi_tQ_b\int
dx_1dx_2\int
db_1db_2b_1b_2\alpha_s(t_8^a)\exp[-S_{B}(t_8^a)-S_V(t_8^a)]
S_t(x_1)\nonumber\\
&\times & C_{8g}(t_8^a)
\phi_{B}(x_1,b_1)
\left[x_2r_V\phi^a_V(x_2)
+x_1\phi^T_V(x_2)+x_2r_V\phi^v_V(x_2)\right]\nonumber\\
&\times &H_8^{(a)}(A_8b_2, B_8b_1,B_8b_2),\nonumber\\
&&\hspace{2cm}
\left(t_8^a=\mbox{max}(A_8,B_8,1/b_1,1/b_2)\right),
\end{eqnarray}
\begin{eqnarray}
M_8^{S(b)}(Q_d)&=& -M_8^{P(b)}(Q_d)\nonumber\\
&=&
-F^{(0)}\xi_tQ_d\int dx_1dx_2\int
db_1db_2b_1b_2\alpha_s(t_8^b)\exp[-S_{B}(t_8^b)-S_V(t_8^b)]
S_t(x_2)\nonumber\\
&\times &C_{8g}(t_8^b)
\phi_{B}(x_1,b_1)
\left[-3x_2r_V\phi^a_V(x_2)
+(2x_2-x_1)\phi^T_V(x_2)-3x_2r_V\phi^v_V(x_2)\right]\nonumber\\
&\times &H_8^{(b)}(A_8b_1,C_8b_1,C_8b_2),\nonumber\\
&&\hspace{2cm}\left(t_8^b=\mbox{max}(A_8,C_8,1/b_1,1/b_2)\right),
\end{eqnarray}
\begin{eqnarray}
M_8^{S(c)}(Q_q)&=&-M_8^{P(c)}(Q_q)\nonumber\\
&=& 
-F^{(0)}\xi_tQ_q\int dx_1dx_2\int
db_1db_2b_1b_2\alpha_s(t_8^c)\exp[-S_{B}(t_8^c)-S_V(t_8^c)]
S_t(x_1)\nonumber\\
&\times &C_{8g}(t_8^c)
\phi_{B}(x_1,b_1)
\left[x_2r_V\phi^a_V(x_2)
-x_1\phi^T_V(x_2)+x_2r_V\phi^v_V(x_2)\right]\nonumber\\
&\times &H_8^{(c)}\left(\sqrt{|A_8^{'2}|}b_2,D_8b_1,D_8b_2\right),\nonumber\\
&&\hspace{2cm}\left(t_8^c=\mbox{max}(\sqrt{|A_8^{'2}|},D_8,1/b_1,1/b_2)\right),
\end{eqnarray}
\begin{eqnarray}
M_8^{S(d)}(Q_q)&=&
-F^{(0)}\xi_tQ_q\int dx_1dx_2\int
db_1db_2b_1b_2\alpha_s(t_8^d)\exp[-S_{B}(t_8^d)-S_V(t_8^d)]
S_t(x_2)\nonumber\\
&\times & C_{8g}(t_8^d)
\phi_{B}(x_1,b_1)
\left[(x_2-x_1+2)\phi^T_V(x_2)
+6x_2r_V\phi^v_V(x_2)\right]\nonumber\\
&\times &H_8^{(d)}\left(\sqrt{|A_8^{'2}|}b_1,E_8b_1,E_8b_2\right),
\end{eqnarray}
\begin{eqnarray}
M_8^{P(d)}(Q_q)&=&
F^{(0)}\xi_tQ_q\int dx_1dx_2\int
db_1db_2b_1b_2\alpha_s(t_8^d)\exp[-S_{B}(t_8^d)-S_V(t_8^d)] 
S_t(x_2)\nonumber\\
&\times &C_{8g}(t_8^d)
\phi_{B}(x_1,b_1)
\left[(x_2-x_1+2)\phi^T_V(x_2)
+6x_2r_V\phi^a_V(x_2)
\right]\nonumber\\
&\times &H_8^{(d)}\left(\sqrt{|A_8^{'2}|}b_1,E_8b_1,E_8b_2\right),\nonumber\\
&&\hspace{2cm}\left(t_8^d=\mbox{max}(\sqrt{|A_8^{'2}|},E_8,1/b_1,1/b_2)\right),
\end{eqnarray}
\begin{eqnarray}
H_8^{(a)}(A_8 b_2,B_8b_1,B_8b_2)
&\equiv &K_0(A_8b_2) 
\Big[
\theta(b_1-b_2)K_0(B_8b_1)I_0(B_8b_2)
+(b_1\leftrightarrow b_2)
\Big],
\end{eqnarray}
\begin{eqnarray}
H_8^{(b)}(A_8 b_1,C_8b_1,C_8b_2)
&\equiv &\frac{i\pi}{2}K_0(A_8b_1)
\Big[
\theta(b_1-b_2)H_0^{(1)}(C_8b_1)J_0(C_8b_2)
+(b_1 \leftrightarrow b_2)
\Big],
\end{eqnarray}
\begin{eqnarray}
H_8^{(c)}(\sqrt{|A_8'^2|}b_2,D_8 b_1,D_8 b_2)
&\equiv &\theta(A_8'^2)\hspace{1mm}
K_0(\sqrt{|A_8'^2|}b_2)
\Big[
\theta(b_1-b_2)K_0(D_8 b_1)I_0(D_8 b_2)
+(b_1\leftrightarrow
b_2)\Big]\nonumber\\
&&\hspace{-2.1cm}+\theta(-A_8'^2)\hspace{1mm}i\frac{\pi}{2}
H_0^{(1)}(\sqrt{|A_8'^2|} b_2) 
\Big[
\theta(b_1-b_2)K_0(D_8 b_1)I_0(D_8 b_2)+(b_1\leftrightarrow
b_2)
\Big],\hspace{1cm}
\end{eqnarray}
\begin{eqnarray}
H_8^{(d)}(\sqrt{|A_8'^2|}b_1,E_8b_1,E_8b_2)
&\equiv &\theta(A_8'^2)\hspace{1mm}
i\frac{\pi}{2}
K_0(\sqrt{|A_8'^2|}b_1)
\left[\theta(b_1-b_2)H_0^{(1)}(E_8b_1)J_0(E_8 b_2)
+(b_1 \leftrightarrow b_2)\right]\nonumber\\
&&\hspace{-3cm}-\theta(-A_8'^2)\hspace{1mm}\left(\frac{\pi}{2}\right)^2
H_0^{(1)}(\sqrt{|A_8'^2}|b_1)
\left[\theta(b_1-b_2)H_0^{(1)}(E_8b_1)J_0(E_8 b_2)
+(b_1 \leftrightarrow b_2)\right],
\end{eqnarray}
\begin{eqnarray}
A_8^2 &=& x_1 x_2 M_B^2,\hspace{8mm}B_8^2=M_B^2(1+x_1),
\hspace{8mm}
C_8^2=M_B^2(1-x_2),\nonumber\\
A_8'^2&=&(x_1-x_2)M_B^2,\hspace{8mm}
D_8^2 =x_1{M_B}^2,\hspace{8mm}E_8^2 =x_2 {M_B}^2.
\end{eqnarray}
Here we define ${Q_q}$ as the electric charge for the quark ${q}$:
${Q_b=Q_d=-1/3}$ and ${Q_u=2/3}$.
Then the decay amplitudes for each decay channels
can be written as follows:
\begin{eqnarray}
M(B^+\to
 \rho^+\gamma)_{8g}^j&=&M_{8g}^{j(a)}(Q_b)+M_{8g}^{j(b)}(Q_d)
+M_{8g}^{j(c)}(Q_u)+M_{8g}^{j(d)}(Q_u),\\
\nonumber\\
M(B^0\to
 \rho^0\gamma)_{8g}^j&=&-\frac{1}{\sqrt{2}}
\left[M_{8g}^{j(a)}(Q_b)+M_{8g}^{j(b)}(Q_d)
+M_{8g}^{j(c)}(Q_d)+M_{8g}^{j(d)}(Q_d)\right],\\
M(B^0\to
 \omega\gamma)_{8g}^j&=&\frac{1}{\sqrt{2}}
\left[M_{8g}^{j(a)}(Q_b)+M_{8g}^{j(b)}(Q_d)
+M_{8g}^{j(c)}(Q_d)+M_{8g}^{j(d)}(Q_d)\right].
\end{eqnarray}

\subsection{Loop contributions}

In this section, we consider the
contributions of diagrams with the
effective operators $O_i$'s inserted in the loop diagram.
${O_1}$ does not contribute because of the color mismatch.
 Penguin
operators $O_{3\sim 6}$ insertion is neglected, 
because they are
small compared with $O_{2}$ insertion in the loop diagram.
 Therefore, we only
consider the tree  $O_{2}$ operator insertion. 
These
diagrams can be separated into two types. One type is that of a
photon emitted from the external quark line (Fig.\ref{loopa}), 
and the
other is that of a photon emitted from the loop quark line
(Fig.\ref{loopb}).

\subsubsection{Contributions of external-quark-line emission}

\begin{figure}
 \begin{center}
\includegraphics[width=4.5cm]{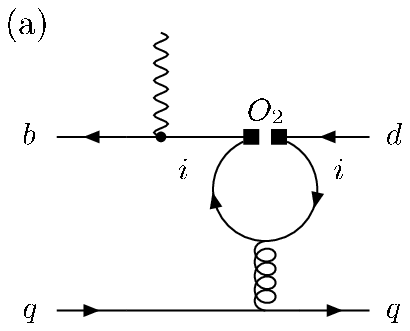}
\hspace{1cm}
\includegraphics[width=4.5cm]{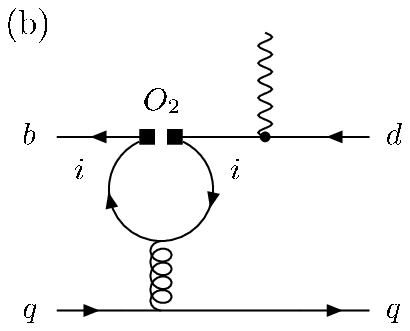}
\vspace{1cm}
\\
\includegraphics[width=4.5cm]{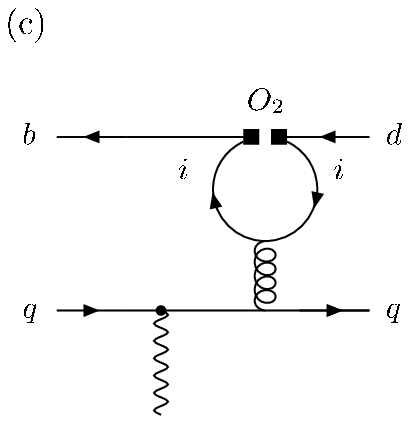}
\hspace{1cm}
\includegraphics[width=4.5cm]{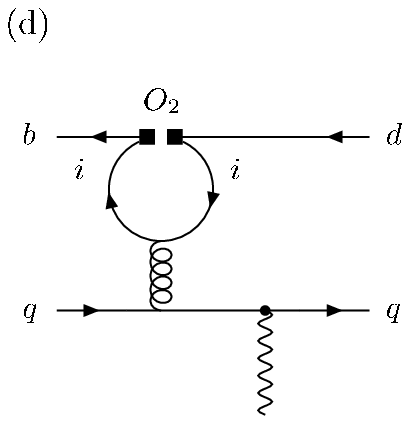}
 \caption{Diagrams in which the operator $O_2$ is inserted 
in the loop, and a
 photon is emitted from the external quark line.
${O_1}$ does not contribute and it can be shown that
${O_{3\sim 6}}$ can be neglected.}
 \label{loopa}
\end{center}
\end{figure}

For the calculation of the diagrams in Fig.\ref{loopa}, one can at
first calculate the effective vertex $\bar{b}\to \bar{d}g$ 
by performing the loop
integration. For the topological structure with  $O_2$ inserted in
the loop diagram of Fig.\ref{loopa}, the effective vertex obtained
with $\overline{\mathrm{MS}}$ scheme is
\begin{eqnarray}
I^{\nu}=\frac{g}{8\pi^2}
\left[\frac{2}{3}-G(m_i^2,k^2,\mu)\right]
\bar{b}T^a_{ij}
 (k^2\gamma^{\nu}-k^{\nu}\not k)(1-\gamma_5)d\; ,\label{bdg}
\\
G(m_i^2,k^2,\mu)= -\int_0^1 dx 4x(1-x)
\mathrm{log}\left[\frac{m_i^2-x(1-x)k^2-i\epsilon}{\mu^2}\right]\;
,
\end{eqnarray}
where $i=u,c$ is the flavor of the loop quark, $k$ is the momentum
of the virtual gluon, and $\nu$ is the Lorentz index of the gluon
field. We can see  that the vertex function
has gauge invariant form.
With the effective vertex given in Eq.(\ref{bdg}), the
contributions of diagrams in Fig.\ref{loopa} can be obtained as
follows:

\begin{eqnarray}
M_{1i}^{S(a)}(Q_b)&=&M_{1i}^{P(a)}(Q_b)\nonumber\\
&=&
\frac{1}{2}F^{(0)}\xi_iQ_b\int dx_1dx_2\int
db_1db_2b_1b_2\alpha_s(t_8^a)\exp[-S_{B}(t_{8}^a)-S_V(t_{8}^a)]
S_t(x_1) \nonumber\\
&\times &C_2(t_8^a)\phi_{B}(x_1,b_1)
x_1x_2r_V\left[\phi^v_V(x_2)
-\phi^a_V(x_2)\right] H_{8}^{(a)}(A_8b_2, B_8b_1,B_8b_2) \nonumber\\
&\times &\left[G(m^2_i,-x_1x_2m_B^2,t_{8}^a)-\frac{2}{3}\right],
\end{eqnarray}
\begin{eqnarray}
M_{1i}^{S(b)}(Q_d)&=&-M_{1i}^{P(b)}(Q_d)\nonumber\\
&=&
-\frac{1}{2}F^{(0)}\xi_iQ_d\int dx_1dx_2\int
db_1db_2b_1b_2\alpha_s(t_{8}^b)\exp[-S_{B}(t_{8}^b)-S_V(t_{8}^b)]
S_t(x_2)\nonumber\\
&&\times C_2(t_8^b)\phi_{B}(x_1,b_1)
\left[3x_1x_2\phi^T_V(x_2)
+x_2^2r_V\{\phi^v_V(x_2)+\phi^a_V(x_2)\}\right]
H_{8}^{(b)}(A_8b_1,C_8b_1, C_8b_2)\nonumber\\
&\times&\left[G(m^2_i,-x_1x_2m_B^2,t_{8}^b)-\frac{2}{3}\right],
\end{eqnarray}
\begin{eqnarray}
M_{1i}^{S(c)}(Q_q)&=& -M_{1i}^{P(c)}(Q_q)\nonumber\\
&=&
\frac{1}{2}F^{(0)}\xi_iQ_q\int dx_1dx_2\int
db_1db_2b_1b_2\alpha_s(t_{8}^c)\exp[-S_{B}(t_{8}^c)-S_V(t_{8}^c)]
S_t(x_1)\nonumber\\
&\times &C_2(t_8^c)\phi_{B}(x_1,b_1)
\left[x_2r_V\{\phi^v_V(x_2)+\phi^a_V(x_2)\}
-x_1\phi^T_V(x_2)\right] 
H_{8}^{(c)}(\sqrt{|A_8^{'2}|}b_2,D_8b_1,D_8b_2)
\nonumber\\
&\times &\left[G\left(m^2_i,(x_2-x_1)m_B^2,t_{8}^c\right)-\frac{2}{3}\right],
\end{eqnarray}
\begin{eqnarray}
M_{1i}^{S(d)}(Q_q)&=&
\frac{1}{2}F^{(0)}\xi_iQ_q\int dx_1dx_2\int
db_1db_2b_1b_2\alpha_s(t_{8}^d) \exp[-S_{B}(t_{8}^d)-S_V(t_{8}^d)]
S_t(x_2)\nonumber\\
&\times &C_2(t_8^d)\phi_{B}(x_1,b_1)
\left[3(x_2-x_1)\phi^T_V(x_2)
+x_2r_V
\{3(1+x_2)\phi^v_V(x_2)-(1-x_2)\phi^a_V(x_2)\}\right]\nonumber\\
&\times &H_{8}^{(d)}(\sqrt{|A_8^{'2}|}b_1, E_8b_1,E_8b_2)
\left[G\left(m^2_i,(x_2-x_1)m_B^2,t_{8}^d\right)-\frac{2}{3}\right],
\end{eqnarray}
\begin{eqnarray}
M_{1i}^{P(d)}(Q_q)&=&
-\frac{1}{2}F^{(0)}\xi_iQ_q\int dx_1dx_2\int
db_1db_2b_1b_2\alpha_s(t_{8}^d)\exp[-S_{B}(t_{8}^d)-S_V(t_{8}^d)]
S_t(x_2)\nonumber\\
&\times &C_2(t_8^d)\phi_{B}(x_1,b_1)
\left[3(x_2-x_1)\phi^T_V(x_2)
-x_2r_V
\{(1-x_2)\phi^v_V(x_2)-3(1+x_2)\phi^a_V(x_2)\}\right]\nonumber\\
&\times &H_{8}^{(d)}(\sqrt{|A_8^{'2}|}b_1, E_8b_1,E_8b_2)
\left[G\left(m^2_i,(x_2-x_1)m_B^2,t_{8}^d\right)-\frac{2}{3}\right].
\end{eqnarray}
Then the decay amplitudes in this case can be expressed as follows:
\begin{eqnarray}
M(B^+\to
 \rho^+\gamma)_{1i}^j&=&M_{1i}^{j(a)}(Q_b)+M_{1i}^{j(b)}(Q_d)
+M_{1i}^{j(c)}(Q_u)+M_{1i}^{j(d)}(Q_u),\\
\nonumber\\
M(B^0\to
 \rho^0\gamma)_{1i}^j&=&-\frac{1}{\sqrt{2}}
\left[M_{1i}^{j(a)}(Q_b)+M_{1i}^{j(b)}(Q_d)
+M_{1i}^{j(c)}(Q_d)+M_{1i}^{j(d)}(Q_d)\right],\\
M(B^0\to
 \omega\gamma)_{1i}^j&=&\frac{1}{\sqrt{2}}
\left[M_{1i}^{j(a)}(Q_b)+M_{1i}^{j(b)}(Q_d)
+M_{1i}^{j(c)}(Q_d)+M_{1i}^{j(d)}(Q_d)\right].
\end{eqnarray}

\subsubsection{Contributions of internal-loop-quark-line emission}

\begin{figure}
\begin{center}
\includegraphics[width=5cm]{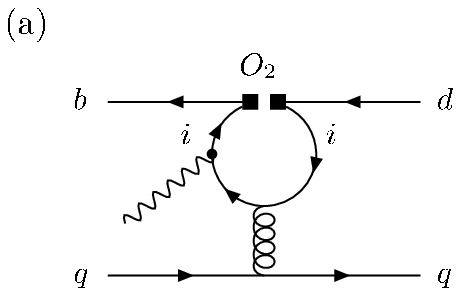}
\hspace{1cm}
\includegraphics[width=5cm]{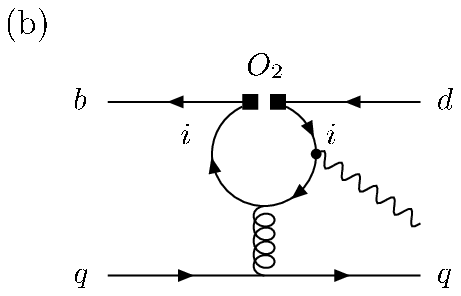}
 \caption{Diagrams in which the operator $O_2$ is
 inserted in the loop, and a
 photon is emitted from the internal loop quark line.}
  \label{loopb}
\end{center}
\end{figure}

The diagrams in which a photon emitted from the internal
loop quark line are shown in Fig.\ref{loopb}. The sum of the
effective vertex $\bar{b}\to \bar{d} \gamma g^*$ in Figs.\ref{loopb}(a) and
\ref{loopb}(b) has been derived in \cite{bdgphoton,bdgphoton2}. The result
can be expressed as
\begin{equation}
I=\bar{d}\gamma^{\rho}(1-\gamma_5)T^a
bI_{\mu\nu\rho}\varepsilon_{\gamma}^{\mu}\varepsilon_{g}^{\nu},
\end{equation}
with the tensor structure given by
\begin{eqnarray}
I_{\mu\nu\rho}
&=&A_4\left[(q\cdot
k)\epsilon_{\mu\nu\rho\sigma}(q-k)^{\sigma}
+\epsilon_{\nu\rho\sigma\tau}q^{\sigma}k^{\tau}k_{\mu}
-\epsilon_{\mu\rho\sigma\tau}q^{\sigma}k^{\tau}q_{\nu}\right]\nonumber\\
&&\hspace{-3mm}+A_5\left[\epsilon_{\mu\rho\sigma\tau}q^{\sigma}k^{\tau}k_{\nu}
     -k^2\epsilon_{\mu\nu\rho\sigma}q^{\sigma}\right],
\end{eqnarray}
and
\begin{eqnarray}
A_4&=&\frac{4ieg}{3\pi^2}\int_0^1dx\int_0^{1-x}dy
\frac{xy}{x(1-x)k^2+
 2xyq\cdot k-m_i^2+i\varepsilon}\; ,\\
A_5&=&-\frac{4ieg}{3\pi^2}\int_0^1dx\int_0^{1-x}dy
\frac{x(1-x)}{x(1-x)k^2+
 2xyq\cdot k-m_i^2+i\varepsilon}\; ,
\end{eqnarray}
where $q$ is the momentum of the photon $q=P_B-P_V$, 
and $k$
is the momentum of the gluon $k=k_2-k_1$.
The result of the amplitudes $M^S$ and $M^P$ contributed by each
diagram in Fig.\ref{loopb} can be expressed as
\begin{eqnarray}
M_{2i}^S&=&
-\frac{4}{3}F^{(0)}\xi_i\int_0^1
dx\int_0^{1-x}dy\int dx_1dx_2\int db_1 b_1\alpha_s(t_{2i})
\exp[-S_{B}(t_{2i}
]\nonumber\\
&\times &C_2(t_{2i})\phi_{B}(x_1,b_1)H_{2i}(b_1A,b_1\sqrt{|B^2|})
\frac{1}{xyx_2M_B^2-m_i^2}\nonumber\\
&\times &\Big[x(1-x)x_2\left(3x_1\phi^T_V(x_2)
+x_2r_V\{\phi^v_V(x_2) +\phi^a_V(x_2)\}\right)\nonumber\\&&
-xyx_2\left((1+2x_1)\phi^T_V(x_2)-r_V\{(1-2x_2)\phi^v_V(x_2)
+\phi^a_V(x_2)\}\right)\Big],
\end{eqnarray}
\begin{eqnarray}
M_{2i}^P&=&
\frac{4}{3}F^{(0)}\xi_i\int_0^1
dx\int_0^{1-x}dy\int dx_1dx_2\int db_1 b_1\alpha_s(t_{2i})
\exp[-S_{B}(t_{2i})
]\nonumber\\
&\times &C_2(t_{2i})\phi_{B}(x_1,b_1)H_{2i}(b_1A,b_1\sqrt{|B^2|})
\frac{1}{xyx_2M_B^2-m_i^2}\nonumber\\
&\times &
\Big[x(1-x)x_2
\left(3x_1\phi^T_V(x_2)+x_2r_V\{\phi^v_V(x_2)
 +\phi^a_V(x_2)\}\right)
\nonumber\\ &&
-xyx_2\left((1+2x_1)\phi^T_V(x_2)-r_V\{(1-2x_2)\phi^a_V(x_2)
+\phi^v_V(x_2)\}\right)\Big],\nonumber\\
&&\hspace{3cm}\left(
t_{2i}=\max (A,\sqrt{|B^2|},1/b_1)\right),
\end{eqnarray}
\begin{eqnarray}
A^2=x_1x_2M_B^2, \hspace{1cm}
B^2=x_1x_2M_B^2 -\frac{y}{1-x}x_2M_B^2+\frac{m_i^2}{x(1-x)},
\end{eqnarray}
\begin{eqnarray}
H_{2i}(b_1A,b_1\sqrt{|B^2|})&\equiv &
K_0(b_1A)-K_0(b_1 \sqrt{|B^2|})\hspace{1.2cm}(B^2 \geq 0),\nonumber\\
&\equiv &K_0(b_1A)-i\frac{\pi}{2}H_0(b_1 \sqrt{|B^2|})\hspace{8mm}(B^2 < 0).
\end{eqnarray}
and the decay amplitudes for each decay modes
are as follows:
\begin{eqnarray}
M(B^+\to\rho^+\gamma)_{2i}^j=M_{2i}^{j},\hspace{8mm}
M(B^0\to\rho^0\gamma)_{2i}^j=-\frac{M_{2i}^{j}}{\sqrt{2}},\hspace{8mm}
M(B^0\to\omega\gamma)_{2i}^j=\frac{M_{2i}^{j}}{\sqrt{2}}.
\end{eqnarray}

\subsection{Annihilation diagram contributions}

Next we consider the annihilation-type diagrams.
They provide the main contribution for
the isospin breaking effects 
in $Br(B^+\to \rho^+\gamma)$ and $2 Br(B^0\to \rho^0\gamma)$.

\subsubsection{Tree annihilation}

\begin{figure}
\begin{center}
\includegraphics[width=4.8cm]{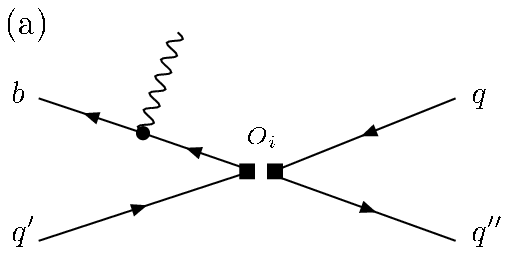}
\hspace{1cm}
\includegraphics[width=4.8cm]{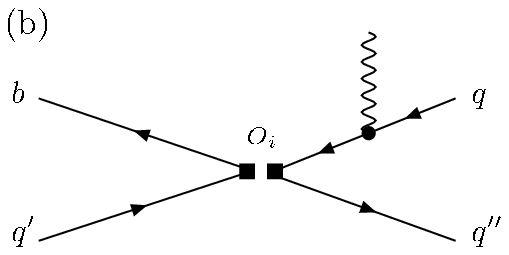}
\vspace{1cm}
\\
\includegraphics[width=4.8cm]{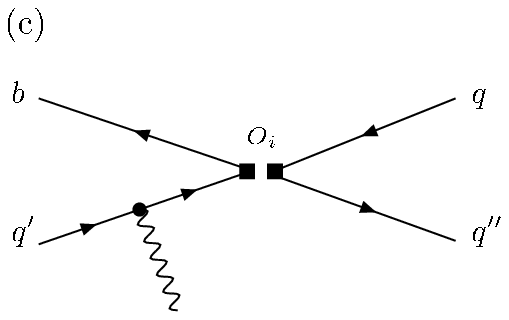}
\hspace{1cm}
\includegraphics[width=4.8cm]{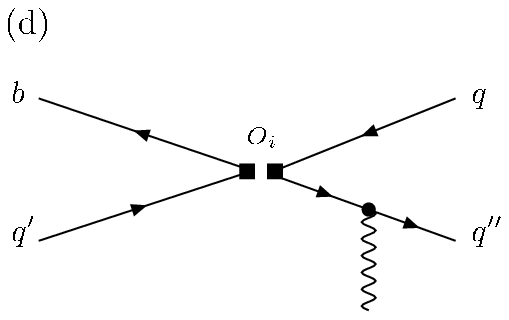}
\end{center}
\caption{Annihilation diagrams in which the operators ${O_1, ~O_2}$ 
are inserted.
  The box denotes operator insertion.} \label{anni1}
\end{figure}

We consider the tree annihilation caused by ${O_1,~O_2}$
operators  shown in Fig.\ref{anni1}. 

In the charged mode, this contribution is color allowed;
on the other hand,
it is color suppressed in the neutral modes.
We define the combinations of the Wilson coefficients as 
\begin{eqnarray}
a_1(t)=C_1(t)+C_2(t)/3,\hspace{1cm}a_2(t)=C_2(t)+C_1(t)/3,
\label{Wil}
\end{eqnarray}
and each decay amplitudes can be given as follows:

\begin{eqnarray}
M_{A_k}^{S(a)}(Q_b)&=&M_{A_k}^{P(a)}(Q_b)\nonumber\\
&=& 
-F^{(0)}\xi_u\frac{3\sqrt{6}Q_bf_V\pi}{4M_B^2}r_V\int dx_1\int
db_1b_1
\exp[-S_{B}(t_{A}^a)]S_t(x_1)\nonumber\\
&\times &a_k(t_{A}^a)\phi_{B}(x_1,b_1)
K_0(b_1A_a),
\hspace{1cm}\left(t_{A}^a=\mbox{max}(A_a,1/b_1)\right),
\end{eqnarray}
\begin{eqnarray}
M_{A_k}^{S(b)}(Q_q)&=&-
F^{(0)}\xi_u\frac{3\sqrt{6}Q_qf_V\pi}{4M_B^2}r_V\int dx_2\int
db_2b_2
\exp[-S_V(t_{A}^b)]S_t(x_2)\nonumber\\
&\times &a_k(t_{A}^b)
i\frac{\pi}{2}H^{(1)}_0(b_2B_a)
\left[x_2\phi^a_V(x_2)+(2-x_2)\phi^v_V(x_2)\right],
\end{eqnarray}
\begin{eqnarray}
M_{A_k}^{P(b)}(Q_q)&=&
F^{(0)}\xi_u\frac{3\sqrt{6}Q_qf_B\pi}{4M_B^2}r_V\int dx_2\int
db_2b_2
\exp[-S_V(t_{A}^b)]S_t(x_2)\nonumber\\
&\times &a_k(t_{A}^b)
i\frac{\pi}{2}H^{(1)}_0(b_2B_a)
\left[(2-x_2)\phi^a_V(x_2)+x_2\phi^v_V(x_2)\right],\nonumber\\
&&\hspace{2cm}\left(t_{A}^b=\mbox{max}(B_a,1/b_2)\right),
\end{eqnarray}
\begin{eqnarray}
M_{A_k}^{S(c)}(Q_{q'})&=& -M_{A_k}^{P(c)}(Q_{q'})\nonumber\\
&=& 
F^{(0)}\xi_u\frac{3\sqrt{6}Q_{q'}f_V\pi}{4M_B^2}r_V\int dx_1\int
db_1b_1\exp[-S_{B}(t_{A}^c)]S_t(x_1)\nonumber\\
&\times &a_k(t_{A}^c)\phi_{B}(x_1,b_1)
K_0(b_1C_1),
\hspace{1cm}\left(t_{A}^c=\mbox{max}(C_a,1/b_1)\right),
\end{eqnarray}
\begin{eqnarray}
M_{A_k}^{S(d)}(Q_{q''})&=&
F^{(0)}\xi_u\frac{3\sqrt{6}Q_{q''}f_B\pi}{4M_B^2}r_V
\int
dx_2\int
db_2b_2\exp[-S_V(t_{A}^d)]S_t(x_2)\nonumber\\
&\times &a_k(t_{A}^d)i\frac{\pi}{2}H^{(1)}_0(b_2D_a)
\left[-(1-x_2)\phi^a_V(x_2)+(1+x_2)\phi^v_V(x_2)\right],
\end{eqnarray}
\begin{eqnarray}
M_{A_k}^{P(d)}(Q_{q''})&=&
F^{(0)}\xi_u\frac{3\sqrt{6}Q_{q''}f_B\pi}{4M_B^2}r_V
\int dx_2\int
db_2b_2
\exp[-S_V(t_{A}^d)]S_t(x_2)\nonumber\\
&\times &a_k(t_{A}^d)i\frac{\pi}{2}H^{(1)}_0(b_2D_a)
\left[(1+x_2)\phi^a_V(x_2)-(1-x_2)\phi^v_V(x_2)\right],\nonumber\\
&&\hspace{2cm}\left(t_{A}^d=\mbox{max}(D_a,1/b_2)\right),
\end{eqnarray}
\begin{eqnarray}
A_a^2=(1+x_1)M_B^2,\hspace{1cm}B_a^2=(1-x_2)M_B^2,\hspace{1cm}C_a^2=x_1M_B^2,
\hspace{1cm}D_a^2=x_2M_B^2.
\end{eqnarray}
Here we use the index ${k}$ in order to express
the Wilson coefficient combination in Eq.(\ref{Wil}).
Then  each decay amplitude can be expressed as follows:
\begin{eqnarray}
M(B^+\to\rho^+\gamma)_{A}^{j}&=&M_{A_2}^{j(a)}(Q_b)+
M_{A_2}^{j(b)}(Q_d)+M_{A_2}^{j(c)}(Q_u)+
M_{A_2}^{j(d)}(Q_u),\\
\nonumber\\
M(B^0\to\rho^0\gamma)_{A}^{j}&=&\frac{1}{\sqrt{2}}
\left[M_{A_1}^{j(a)}(Q_b)
+M_{A_1}^{j(b)}(Q_u)+M_{A_1}^{j(c)}(Q_d)+M_{A_1}^{j(d)}(Q_u)\right],
\\
M(B^0\to\omega\gamma)_{A}^{j}&=&
\frac{1}{\sqrt{2}}
\left[M_{A_1}^{j(a)}(Q_b)
+M_{A_1}^{j(b)}(Q_u)+M_{A_1}^{j(c)}(Q_d)
+M_{A_1}^{j(d)}(Q_u)\right].
\end{eqnarray}

\subsubsection{QCD penguin annihilation}

\begin{figure}
\begin{center}
\includegraphics[width=4.8cm]{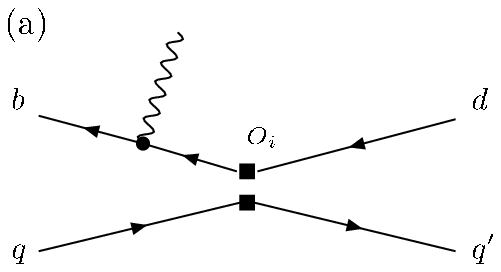}
\hspace{1cm}
\includegraphics[width=4.8cm]{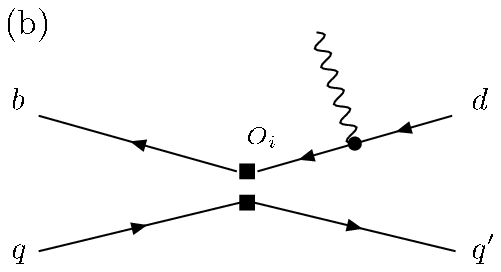}
\vspace{1cm}
\\
\includegraphics[width=4.8cm]{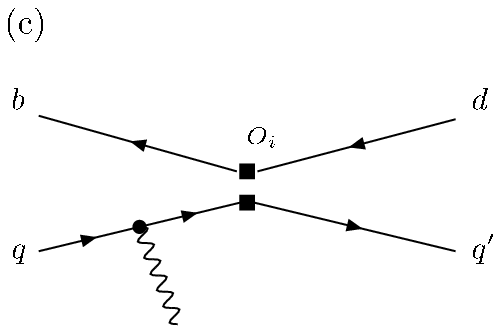}
\hspace{1cm}
\includegraphics[width=4.8cm]{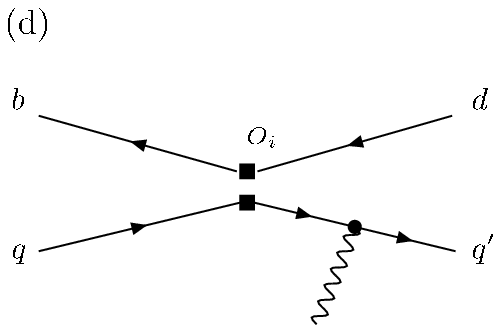}
\end{center}
\caption{The other type of annihilation diagrams with operator
insertion.} \label{anni3}
\end{figure}

Next we consider the QCD penguin annihilation contributions.
There are two types of the
annihilation diagrams in which the operators $O_i$'s are inserted.
One type is shown in Fig.\ref{anni2} and the other
is in Fig.\ref{anni3}.

First we consider the Fig.\ref{anni2} contributions.
Here we also define the combinations of the
Wilson coefficients as
\begin{eqnarray}
&&a_3(t)=C_3(t)+C_4(t)/3,\hspace{1cm}a_4(t)=C_4(t)+C_3(t)/3,\nonumber\\
&&a_5(t)=C_5(t)+C_6(t)/3,\hspace{1cm}a_6(t)=C_6(t)+C_5(t)/3.
\label{Wil2}
\end{eqnarray}
For ${(V-A)(V-A)}$ operators, the results are as follows:

\begin{figure}
\begin{center}
\includegraphics[width=4.8cm]{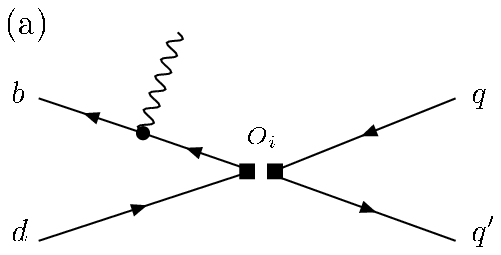}
\hspace{1cm}
\includegraphics[width=4.8cm]{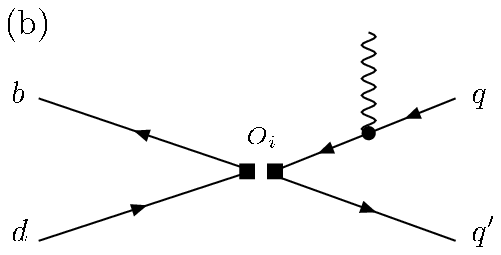}
\vspace{1cm}
\\
\includegraphics[width=4.8cm]{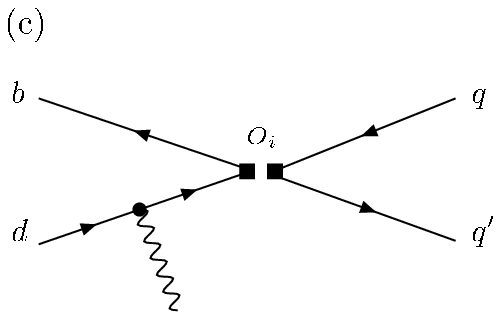}
\hspace{1cm}
\includegraphics[width=4.8cm]{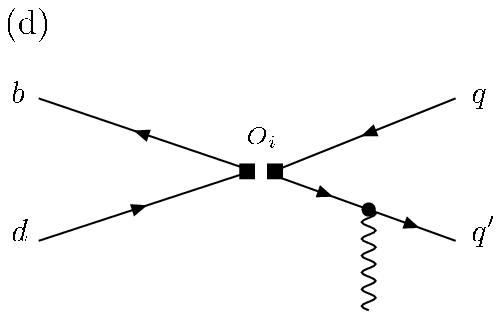}
\end{center}
\caption{Annihilation diagrams in which the QCD penguin operators are
inserted.
  The box denotes operator insertion.} \label{anni2}
\end{figure}

\begin{eqnarray}
M_{A1_k}^{S(a)^-}(Q_b)&=&M_{A1_k}^{P(a)^-}(Q_b)\nonumber\\
&=& 
F^{(0)}\xi_t\frac{3\sqrt{6}Q_bf_V\pi}{4M_B^2}r_V\int dx_1\int
db_1b_1
\exp[-S_{B}(t_{A1}^a)]S_t(x_1)\nonumber\\
&\times &a_k(t_{A1}^a)\phi_{B}(x_1,b_1)
K_0(b_1A_a),
\hspace{1cm}\left(t_{A1}^a=\mbox{max}(A_a,1/b_1)\right),
\end{eqnarray}
\begin{eqnarray}
M_{A1_k}^{S(b)^-}(Q_q)&=&
F^{(0)}\xi_t\frac{3\sqrt{6}Q_qf_V\pi}{4M_B^2}r_V\int dx_2\int
db_2b_2
\exp[-S_V(t_{A1}^b)]S_t(x_2)\nonumber\\
&\times &a_k(t_{A1}^b)
i\frac{\pi}{2}H^{(1)}_0(b_2B_a)
\left[x_2\phi^a_V(x_2)+(2-x_2)\phi^v_V(x_2)\right],
\end{eqnarray}
\begin{eqnarray}
M_{A1_k}^{P(b)^-}(Q_q)&=&-
F^{(0)}\xi_t\frac{3\sqrt{6}Q_qf_B\pi}{4M_B^2}r_V\int dx_2\int
db_2b_2
\exp[-S_V(t_{A1}^b)]S_t(x_2)\nonumber\\
&\times &a_k(t_{A1}^b)i\frac{\pi}{2}H^{(1)}_0(b_2B_a)
\left[(2-x_2)\phi^a_V(x_2)+x_2\phi^v_V(x_2)\right],\nonumber\\
&&\hspace{2cm}\left(t_{A1}^b=\mbox{max}(B_a,1/b_2)\right),
\end{eqnarray}
\begin{eqnarray}
M_{A1_k}^{S(c)^-}(Q_{d})&=& -M_{A1_k}^{P(c)^-}(Q_{d})\nonumber\\
&=&- 
F^{(0)}\xi_t\frac{3\sqrt{6}Q_{d}f_V\pi}{4M_B^2}r_V\int dx_1\int
db_1b_1\exp[-S_{B}(t_{A1}^c)]S_t(x_1)\nonumber\\
&\times &a_k(t_{A1}^c)\phi_{B}(x_1,b_1)
K_0(b_1C_1),
\hspace{1cm}\left(t_{A1}^c=\mbox{max}(C_a,1/b_1)\right),
\end{eqnarray}
\begin{eqnarray}
M_{A1_k}^{S(d)^-}(Q_{q'})&=&-
F^{(0)}\xi_t\frac{3\sqrt{6}Q_{q'}f_B\pi}{4M_B^2}r_V
\int
dx_2\int
db_2b_2\exp[-S_V(t_{A1}^d)]S_t(x_2)\nonumber\\
&\times &a_k(t_{A1}^d)i\frac{\pi}{2}H^{(1)}_0(b_2D_a)
\left[-(1-x_2)\phi^a_V(x_2)+(1+x_2)\phi^v_V(x_2)\right],
\end{eqnarray}
\begin{eqnarray}
M_{A1_k}^{P(d)^-}(Q_{q'})&=&-
F^{(0)}\xi_t\frac{3\sqrt{6}Q_{q'}f_B\pi}{4M_B^2}r_V
\int dx_2\int
db_2b_2
\exp[-S_V(t_{A1}^d)]S_t(x_2)\nonumber\\
&\times &a_k(t_{A1}^d)i\frac{\pi}{2}H^{(1)}_0(b_2D_a)
\left[(1+x_2)\phi^a_V(x_2)-(1-x_2)\phi^v_V(x_2)\right]\nonumber\\
&&\hspace{2cm}\left(t_{A1}^d=\mbox{max}(D_a,1/b_2)\right),
\end{eqnarray}
\begin{eqnarray}
A_a^2=(1+x_1)M_B^2,\hspace{1cm}B_a^2=(1-x_2)M_B^2,\hspace{1cm}C_a^2=x_1M_B^2,
\hspace{1cm}D_a^2=x_2M_B^2.
\end{eqnarray}
Here we use the index ${k}$ in order to express
the Wilson coefficient combination in Eq.(\ref{Wil2}),
and upper index ``-'' means the
${(V-A)(V-A)}$ vertex structure.
The total amplitudes in Fig.\ref{anni2} 
 contribute only to the neutral decay modes and they
are color suppressed contributions, thus they
are given as follows:

\begin{eqnarray}
M(B^0\to\rho^0\gamma)_{A1}^{j^-}&=&
\frac{1}{\sqrt{2}}
\Big[\left(M_{A1_3}^{j(b)^-}(Q_u)-M_{A1_3}^{j(b)^-}(Q_d)\right)
+
\left(M_{A1_3}^{j(d)^-}(Q_u)-M_{A1_3}^{j(d)^-}(Q_d)\right)\Big],\hspace{8mm}\\
M(B^0\to\omega\gamma)_{A1}^{j^-}&=&
\frac{1}{\sqrt{2}}
\Big{[} 2M_{A1_3}^{j(a)^-}(Q_b)
+\left\{M_{A1_3}^{j(b)^-}(Q_u)+M_{A1_3}^{j(b)^-}(Q_d)\right\}\nonumber\\
&&\hspace{20mm}
+2M_{A1_3}^{j(c)^-}(Q_d)
+\left\{M_{A1_3}^{j(d)^-}(Q_u)+M_{A1_3}^{j(d)^-}(Q_d)\right\}\Big{]}.\hspace{5mm}
\end{eqnarray}

Amplitudes with ${(V-A)(V+A)}$ operators
can be related to those with
${(V-A)(V-A)}$ amplitudes as
\begin{eqnarray}
M_{A1}^{S(a)^+}(Q_b)&=&M_{A1}^{S(a)^-}(Q_b),\hspace{1cm}
M_{A1}^{P(a)^+}(Q_b)=M_{A1}^{P(a)^-}(Q_b),\\
M_{A1}^{S(b)^+}(Q_q)&=&M_{A1}^{S(b)^-}(Q_q),\hspace{1cm}
M_{A1}^{P(b)^+}(Q_q)=-M_{A1}^{P(b)^-}(Q_q),\\
M_{A1}^{S(c)^+}(Q_{q'})&=&M_{A1}^{S(c)^-}(Q_{q'}),\hspace{1cm}
M_{A1}^{P(c)^+}(Q_{q'})=M_{A1}^{P(c)^-}(Q_{q'}),\\
M_{A1}^{S(d)^+}(Q_{q''})&=&M_{A1}^{S(d)^-}(Q_{q''}),\hspace{1cm}
M_{A1}^{P(d)^+}(Q_{q''})=-M_{A1}^{P(d)^-}(Q_{q''}),
\end{eqnarray}
where ``+'' expresses the ${(V-A)(V+A)}$ vertex structure.
The decay amplitudes caused by the  ${(V-A)(V+A)}$ vertex
exist only in the neutral modes and
can be expressed as follows:
\begin{eqnarray}
M(B^0\to\rho^0\gamma)_{A1}^{j^+}&=&
\frac{1}{\sqrt{2}}\left[
\Big{\{ }M_{A1_5}^{j(b)^+}(Q_u)-M_{A1_5}^{j(b)^+}(Q_d)\Big{\}}
+\Big{\{ }M_{A1_5}^{j(d)^+}(Q_u)-M_{A1_5}^{j(d)^+}(Q_d)\Big{\}}
\right],\hspace{5mm}\\
M(B^0\to\omega\gamma)_{A1}^{j^+}&=&
\frac{1}{\sqrt{2}}
\Big[
2M_{A1_5}^{j(a)^+}(Q_b)+\Big{\{ }M_{A1_5}^{j(b)^+}(Q_u)+M_{A1_5}^{j(b)^+}(Q_d)\Big{\}}\nonumber\\
&&\hspace{2cm}+
2M_{A1_5}^{j(c)^+}(Q_d)+\Big{\{ }M_{A1_5}^{j(d)^+}(Q_u)+M_{A1_5}^{j(d)^+}(Q_d)\Big{\}}
\Big].
\end{eqnarray}

Next we consider the type two diagrams shown in Fig.\ref{anni3}.
For ${(V-A)(V-A)}$ operators
inserted in these diagrams, the results $M_{A2}^{(S,P)^-}$
are the same as
$M_{A1}^{(S,P)^-}$, 
\begin{eqnarray}
M(B^+\to\rho^+\gamma)_{A2}^{j^-}&=&M_{A2_4}^{j(a)^-}(Q_b)+
M_{A2_4}^{j(b)^-}(Q_d)+M_{A2_4}^{j(c)^-}(Q_u)+
M_{A2_4}^{j(d)^-}(Q_u),\\
\nonumber\\
M(B^0\to\rho^0\gamma)_{A2}^{j^-}&=&-\frac{1}{\sqrt{2}}
\left[M_{A2_4}^{j(a)^-}(Q_b)+
M_{A2_4}^{j(b)^-}(Q_d)+M_{A2_4}^{j(c)^-}(Q_d)+
M_{A2_4}^{j(d)^-}(Q_d)\right],\\
M(B^0\to\omega\gamma)_{A2}^{j^-}&=&\frac{1}{\sqrt{2}}
\left[M_{A2_4}^{j(a)^-}(Q_b)+
M_{A2_4}^{j(b)^-}(Q_d)+M_{A2_4}^{j(c)^-}(Q_d)+
M_{A2_4}^{j(d)^-}(Q_d)\right].
\end{eqnarray}
On the other hand, for ${(V-A)(V+A)}$ operators, 
the results are
\begin{eqnarray}
M_{A2_k}^{S(a)^+}(Q_b)= M_{A2_k}^{P(a)^+}(Q_b)=0 ,
\end{eqnarray}
\begin{eqnarray}
M_{A2_k}^{S(b)^+}(Q_d)&=&-M_{A2_k}^{S(b)^+}(Q_d)\nonumber\\
&=&
F^{(0)}\xi_p\frac{3\sqrt{6}Q_df_B\pi}{2M_B^2}
\int dx_2\int
db_2b_2
\exp[-S_V(t_{A1}^b)]S_t(x_2)\nonumber\\
&\times &a_k(t_{A1}^b)\phi^T_V(x_2) 
i\frac{\pi}{2}H^{(1)}_0(b_2B_a),
\end{eqnarray}
\begin{eqnarray}
M_{A2_k}^{S(c)^+}(Q_q)=M_{A2_k}^{P(c)^+}(Q_q)=0,
\end{eqnarray}
\begin{eqnarray}
M_{A2_k}^{S(d)^+}(Q_{q'})&=&-M_{A2_k}^{S(d)^+}(Q_{q'})\nonumber\\
&=&
F^{(0)}\xi_p\frac{3\sqrt{6}Q_{q'}f_B\pi}{2M_B^2}\int dx_2\int
db_2b_2
\exp[-S_V(t_{A1}^d)]S_t(x_2) \nonumber\\
&\times &a_k(t_{A1}^d)\phi^T_V(x_2) 
i\frac{\pi}{2}H^{(1)}_0(b_2D_a),
\end{eqnarray}
then the amplitudes of this type with ${(V-A)(V+A)}$
vertex become as follows:
\begin{eqnarray}
M(B^+\to\rho^+\gamma)_{A2}^{j^+}&=&M_{A2_6}^{j(a)^+}(Q_b)+
M_{A2_6}^{j(b)^+}(Q_d)+M_{A2_6}^{j(c)^+}(Q_u)+
M_{A2_6}^{j(d)^+}(Q_u),\\
\nonumber\\
M(B^0\to\rho^0\gamma)_{A2}^{j^+}&=&-\frac{1}{\sqrt{2}}
\left[M_{A2_6}^{j(a)^+}(Q_b)+
M_{A2_6}^{j(b)^+}(Q_d)+M_{A2_6}^{j(c)^+}(Q_d)+
M_{A2_6}^{j(d)^+}(Q_d)\right],\\
M(B^0\to\omega\gamma)_{A2}^{j^+}&=&\frac{1}{\sqrt{2}}
\left[M_{A2_6}^{j(a)^+}(Q_b)+
M_{A2_6}^{j(b)^+}(Q_d)+M_{A2_6}^{j(c)^+}(Q_d)+
M_{A2_6}^{j(d)^+}(Q_d)\right].
\end{eqnarray}
In these type two cases, they are all color allowed decay modes.

\subsection{The final decay amplitudes $M^S$ and $M^P$}

Finally, 
we summarize the amplitudes $M^j$'s, $j=S,P$ for each decay mode: 
\begin{eqnarray}
M^j(B^+\to\rho^+\gamma)&=&M(B^+\to \rho^+\gamma)^j_{7\gamma}
+M(B^+\to \rho^+\gamma)^j_{8g}+M(B^+\to \rho^+\gamma)^j_{1i}
\nonumber\\
&&\hspace{-3mm}
+M(B^+\to \rho^+\gamma)^j_{2i}
+M(B^+\to \rho^+\gamma)^{j}_{A}
+M(B^+\to \rho^+\gamma)^{j-}_{A2}\nonumber\\
&&\hspace{-3mm}
+M(B^+\to \rho^+\gamma)^{j+}_{A2}~,
\end{eqnarray}
\begin{eqnarray}
M^j(B^0\to\rho^0\gamma)&=&M(B^0\to \rho^0\gamma)^j_{7\gamma}
+M(B^0\to \rho^0\gamma)^j_{8g}+M(B^0\to \rho^0\gamma)^j_{1i}\nonumber\\
&&\hspace{-3mm}
+M(B^0\to \rho^0\gamma)^j_{2i}
+M(B^0\to \rho^0\gamma)^{j}_{A}
+M(B^0\to \rho^0\gamma)^{j-}_{A1}\nonumber\\
&&\hspace{-3mm}
+M(B^0\to \rho^0\gamma)^{j+}_{A1}
+M(B^0\to \rho^0\gamma)^{j-}_{A2}
+M(B^0\to \rho^0\gamma)^{j+}_{A2}~,
\end{eqnarray}
\begin{eqnarray}
M^j(B^0\to\omega\gamma)&=&M(B^0\to \omega\gamma)^j_{7\gamma}
+M(B^0\to \omega\gamma)^j_{8g}
+M(B^0\to \omega\gamma)^j_{1i}
\nonumber\\
&&\hspace{-3mm}
+M(B^0\to \omega\gamma)^j_{2i}
+M(B^0\to \omega\gamma)^{j}_{A}
+M(B^0\to \omega\gamma)^{j-}_{A1}\nonumber\\
&&\hspace{-3mm}
+M(B^0\to \omega\gamma)^{j+}_{A1}
+M(B^0\to \omega\gamma)^{j-}_{A2}
+M(B^0\to \omega\gamma)^{j+}_{A2}~.
\end{eqnarray}

\section{Numerical analysis and discussions}
\label{numerical}
In our numerical calculations, the choice of the input parameters is
summarized in Tab.\ref{parameter},
 where $\lambda$, $A$, $\rho$ and $\eta$ are CKM
parameters in Wolfenstein parametrization \cite{wolfenstein}, and
$\bar{\rho}=\rho (1-\frac{1}{2}\lambda^2)$, $\bar{\eta}=\eta
(1-\frac{1}{2}\lambda^2)$. Their values can be found in PDG
\cite{pdg}.
The numerical results for each decay amplitudes ${M_i^j}$
in the ${B^0\to \rho^0\gamma}$ (Tab.\ref{rho0}),
${B^0\to \omega\gamma}$ (Tab.\ref{omega}),
and ${B^+\to \rho^+\gamma}$ (Tab.\ref{chargedrho}) in
unit of ${10^{-6}\mbox{GeV}^{-2}}$ are as follows.

When we estimate the physical quantities like
branching ratio, direct CP asymmetry, and isospin breaking
effect,
we take into account the following theoretical errors.
The detailed discussions for the errors are in \cite{Keum:2004is}.
First, we change the input parameters;
the decay constants and
${\omega_B}$ in the
$B$ meson wave function, and we regard
the 15\% error in each cases
at the amplitude level.
This generates the theoretical
error for the physical quantities 
about 40\% in the branching ratio, 5\%
in the direct CP asymmetry,
and 30\% in the isospin breaking.

Second, we estimate that the higher order effects
in perturbation expansion to be about 15\% error
in the amplitude.
This leads to
about 30\% in the branching ratio  and in the
isospin breaking.
Here, the cancellation of 
the higher order effects can occur
by taking the ratio of the decay width in
the direct CP asymmetry.
Then we can neglect these uncertainties for the CP asymmetry.

\begin{table}
\begin{center}
\begin{tabular}{c}\hline\hline
CKM parameters and QCD constant\\
 $\begin{array}{ccccc}
 \lambda & A &\bar{\rho}& \bar{\eta}&\Lambda_{\overline{\mathrm{MS}}}^{(f=4)}\\
 0.2196& 0.819&0.20\pm 0.09& 0.33\pm 0.05& 250~ \mbox{MeV}
 \end{array}$\\
 \hline
 Meson decay constants\\
 $\begin{array}{ccccc}
   f_B & f_{\rho} &f_{\rho}^T & f_{\omega}
  &f_{\omega}^T\\
  190~\mbox{MeV}& 220~\mbox{MeV} & 160~\mbox{MeV}& 195~\mbox{MeV} &
  160~\mbox{MeV}
 \end{array}$ \\
 \hline
  Masses\\
 $\begin{array}{ccccc}
  M_W&  M_B  & M_{\rho} & M_{\omega}&m_c\\
   80.41~\mbox{GeV} & 5.28~\mbox{GeV} & 0.77~\mbox{GeV} &
    0.78~\mbox{GeV} 
&1.2~\mbox{MeV}
 \end{array}$ \\
 \hline
 $B$ meson life time \\
 $ \begin{array}{cc}
  \tau_{B^0}& \tau_{B^{\pm}} \\
   1.542~\mbox{ps} & 1.674~\mbox{ps}
 \end{array}$\\ \hline\hline
\end{tabular}
\end{center}
\caption{Summary of input parameters}
\label{parameter}
\end{table}

Third, the error due to the CKM parameter uncertainties
${\bar{\rho}}$ and ${\bar{\eta}}$ generates
about 30\% error in the branching ratios
and direct CP asymmetry,
and $100$\% error in the isospin breaking effects.
We can see that the uncertainty which comes from 
the CKM parameters are large compared to 
${B\to K^*\gamma}$ decay modes \cite{Keum:2004is}.
The reason for it is that the all CKM matrix elements
which concern ${B\to \rho (\omega)\gamma}$ decay
(${V_{tb}^*V_{td}}$, ${V_{cb}^*V_{cd}}$, ${V_{ub}^*V_{ud}}$)
are comparable, and the three angles of the KM unitarity triangle
are sizable: the situation is different from in ${B\to K^*\gamma}$
decay.
The conditions mentioned above make the
uncertainty from the CKM parameters large.

In the end, we also take into account 
the uncertainties from ${u}$ quark loop contributions
like Figs.\ref{loopa} and \ref{loopb}.
We guess that the nonperturbative effects in the
${u}$ quark loop might lead to large
hadronic uncertainties.
So, we introduce the 100\% theoretical error
at the amplitude level.
This theoretical uncertainty leads to
small uncertainties (about 2\%) for
the branching ratio,
${80}$\% errors for the CP asymmetry,
and about 3\% errors for the isospin breaking effects. 

\begin{center}
\begin{table}[htbp]
\hspace{-1.5cm}
\begin{tabular}{c c c|| c c c}
\hline
&${M_i^S/F^{(0)}\xi_q}$&&&${M_i^P/F^{(0)}\xi_q}$&\\
\hline
\hline
${M_{7\gamma}^S/F^{(0)}\xi_t}$&
${M_{8g}^S/F^{(0)}\xi_t}$&
${\sum_{i=1,2}M_{Ai}^S/F^{(0)}\xi_t}$&${M_{7\gamma}^P/F^{(0)}\xi_t}$&
${M_{8g}^P/F^{(0)}\xi_t}$&${\sum_{i=1,2}M_{Ai}^P/F^{(0)}\xi_t}$\\
\hline
 -172.12 &-0.44-1.19i& -7.76 - 3.45 i&
172.12&0.48+1.18i &7.55+ 3.47 i\\
\hline
${M_{1c}^S/F^{(0)}\xi_c}$&
${M_{2c}^S/F^{(0)}\xi_c }$&&
${M_{1c}^P/F^{(0)}\xi_c}$&
${M_{2c}^P/F^{(0)}\xi_c}$&\\
\hline
0.39+1.01i&-1.21+8.84i &&
-0.08-0.97i &0.42-6.28i& \\
\hline
 ${M_{1u}^S/F^{(0)}\xi_u}$&
${M_{2u}^S/F^{(0)}\xi_u}$&${M_A^S/F^{(0)}\xi_u}$
&${M_{1u}^P/F^{(0)}\xi_u}$&
${M_{2u}^P/F^{(0)}\xi_u}$&${M_A^P/F^{(0)}\xi_u}$\\
\hline
 1.35+2.06i &-1.11-27.76i &1.14-0.01i &
-1.13-1.98i &-0.73+28.15i &-2.47+0.17i\\
\hline
\end{tabular}
\caption{The numerical results for ${B^0\to \rho^0\gamma}$ decay
at ${\bar{\rho}=0.20}$, ${\bar{\eta}=0.33}$, ${\omega_B=0.40}$GeV.}
\label{rho0}
\end{table}

\begin{table}[htbp]
\hspace{-1.5cm}
\begin{tabular}{ c c c|| c c c}
\hline
&${M_i^S/F^{(0)}\xi_q}$&&&${M_i^P/F^{(0)}\xi_q}$&\\
\hline
\hline
${M_{7\gamma}^S/F^{(0)}\xi_t}$&
${M_{8g}^S/F^{(0)}\xi_t}$&
${\sum_{i=1,2}M_{Ai}^S/F^{(0)}\xi_t}$&${M_{7\gamma}^P/F^{(0)}\xi_t}$&
${M_{8g}^P/F^{(0)}\xi_t}$&${\sum_{i=1,2}M_{Ai}^P/F^{(0)}\xi_t}$\\
\hline
 161.59 &0.44+1.21i &7.73 + 3.45 i&
-161.59 &-0.47-1.18i &-7.71 - 3.42 i\\
\hline
 ${M_{1c}^S/F^{(0)}\xi_c}$&
${M_{2c}^S/F^{(0)}\xi_c}$&
&${M_{1c}^P/F^{(0)}\xi_c}$&
${M_{2c}^P/F^{(0)}\xi_c}$&\\
\hline
-0.46-0.99i &1.21-8.58i &&0.12+0.99i &
-0.36+6.35i &\\
\hline
${M_{1u}^S/F^{(0)}\xi_u}$&
${M_{2u}^S/F^{(0)}\xi_u}$&${M_A^S/F^{(0)}\xi_u}$
&${M_{1u}^P/F^{(0)}\xi_u}$&
${M_{2u}^P/F^{(0)}\xi_u}$&${M_A^P/F^{(0)}\xi_u}$\\
\hline
-1.03-2.10i &1.64+ 27.00i &1.04-0.01i&
1.06+2.10i &0.05-27.45i &-2.25+0.12i\\
\hline
\end{tabular}
\caption{The numerical results for ${B^0\to \omega\gamma}$ decay
at ${\bar{\rho}=0.20}$, ${\bar{\eta}=0.33}$, ${\omega_B=0.40}$GeV.}
\label{omega}
\end{table}

\begin{table}[htbp]
\hspace{-1.5cm}
\begin{tabular}{ c c c|| c c c}
\hline
&${M_i^S/F^{(0)}\xi_q}$&&&${M_i^P/F^{(0)}\xi_q}$&\\
\hline
\hline
${M_{7\gamma}^S/F^{(0)}\xi_t}$&
${M_{8g}^S/F^{(0)}\xi_t}$&
${\sum_{i=1,2}M_{Ai}^S/F^{(0)}\xi_t}$&${M_{7\gamma}^P/F^{(0)}\xi_t}$&
${M_{8g}^P/F^{(0)}\xi_t}$&${\sum_{i=1,2}M_{Ai}^P/F^{(0)}\xi_t}$\\
\hline
243.72 &4.76-3.12i &-4.61 - 2.73 i&
-243.72 &-4.56+3.15i & 4.08 + 2.42 i\\
\hline
${M_{1c}^S/F^{(0)}\xi_c}$&
${M_{2c}^S/F^{(0)}\xi_c}$&
&${M_{1c}^P/F^{(0)}\xi_c}$&
${M_{2c}^P/F^{(0)}\xi_c}$&\\
\hline
-0.74+2.70i &1.70-15.36i &&
1.51-3.02i &-0.48+ 11.23i &\\
\hline
${M_{1u}^S/F^{(0)}\xi_u}$&
${M_{2u}^S/F^{(0)}\xi_u}$&${M_A^S/F^{(0)}\xi_u}$
&${M_{1u}^P/F^{(0)}\xi_u}$&
${M_{2u}^P/F^{(0)}\xi_u}$&${M_A^P/F^{(0)}\xi_u}$\\
\hline
-2.39+ 6.05i &1.70+39.28i &37.28-7.90i&
2.65-5.19i &0.99-39.80i &-55.47-0.37i\\
\hline
\end{tabular}
\caption{The numerical results for ${B^+\to \rho^+\gamma}$ decay
at ${\bar{\rho}=0.20}$, ${\bar{\eta}=0.33}$, ${\omega_B=0.40}$GeV.}
\label{chargedrho}
\end{table}
\end{center}

Then the total theoretical error for  each physical quantity
becomes as about 60\% in the branching ratio,
85\% in the CP asymmetry, and 100\% in the isospin breaking effects.

With the amplitudes $M^S$ and $M^P$ defined in Eq.(\ref{MSP}), the
total decay rate of $B\to \rho(\omega)\gamma$ is given by
\begin{equation}
\Gamma =\frac{|M^S|^2+|M^P|^2}{8\pi M_B}\; ,
\end{equation}
and the relevant decay branching ratio is defined to be
\begin{equation}
Br=\frac{\tau_B}{\hbar}\Gamma\; ,
\end{equation}
where $\tau_B$ is the mean lifetime of the $B$ meson.
The branching ratios for neutral and charged modes
are defined as
\begin{eqnarray}
Br(B^{\pm}\to \rho^{\pm}\gamma)&=&\frac{1}{2}
\left[
Br(B^{+}\to \rho^{+}\gamma)+Br(B^{-}\to \rho^{-}\gamma)
\right],\hspace{7mm}
\\
Br(B^{0}\to \rho^{0}\gamma)&=&\frac{1}{2}
\left[
Br(B^{0}\to \rho^{0}\gamma)+Br(\bar{B}^0\to \rho^{0}\gamma)
\right],\hspace{7mm}
\end{eqnarray}
and its' predicted values become as 
\begin{eqnarray}
Br(B^{0}\to \rho^{0}\gamma)=(1.2\pm 0.7)\times 10^{-6},\label{eq2}\\
Br(B^{0}\to \omega\gamma)=(1.1\pm 0.6)\times 10^{-6},\label{eq3}\\
Br(B^{\pm}\to \rho^{\pm}\gamma)=(2.5\pm 1.5)\times
 10^{-6}\label{eq1}.
\end{eqnarray}

The direct CP asymmetry is defined by
\begin{equation}
A_{cp}(B^{\pm} \to \rho^{\pm}\gamma)=\frac{\Gamma(B^-\to \rho^-\gamma)
-\Gamma(B^+\to \rho^+\gamma)}{\Gamma(B^-\to \rho^-\gamma)
+\Gamma(B^+\to \rho^+\gamma)}
\end{equation}
for charged $B$ meson decays, and
\begin{eqnarray}
A_{cp}(B^0\to \rho^0(\omega)\gamma)=
\frac{\Gamma(\bar{B}^0\to \rho^0(\omega)\gamma)
-\Gamma(B^0\to \rho^0(\omega)\gamma)}{\Gamma(\bar{B}^0\to \rho^0(\omega)
 \gamma)
+\Gamma(B^0\to \rho^0(\omega)\gamma)}\nonumber\\
\end{eqnarray}
for neutral $B$ meson decays.
The numerical results for these CP asymmetries in $B\to\rho\gamma$ and
$\omega\gamma$ are as follows:
\begin{eqnarray}
A_{cp}(B^{0}\to \rho^{0}\gamma)&=&(17.6\pm 15.0)\%\\
A_{cp}(B^{0}\to \omega\gamma)&=&(17.9\pm 15.2)\%\\
A_{cp}(B^{\pm}\to \rho^{\pm}\gamma)&=&(17.7\pm 15.0)\%.
\end{eqnarray}

Next we discuss the isospin breaking effect in $B\to\rho\gamma$
decay. The isospin relation requires that the branching ratio of
$B^+\to\rho^+\gamma$ is two times of $B^0\to\rho^0\gamma$. However,
the contribution of the annihilation diagrams can violate this isospin
relation.
We can define the isospin breaking
parameter as
\begin{eqnarray}
\Delta_{0+} (B\to\rho\gamma)&=&
\frac{\Gamma(B^+\to\rho^+\gamma)}{2\Gamma(B^0\to\rho^0\gamma)}-1,
\\
\Delta_{0-} (B\to\rho\gamma)&=&
\frac{\Gamma(B^-\to\rho^-\gamma)}{2\Gamma(\bar{B}^0\to\rho^0\gamma)}-1,\\
\Delta(\rho\gamma)&=&\frac{\Delta_{0+} +\Delta_{0-}}{2}.
\label{isospin}
\end{eqnarray}
If isospin relation is maintained, $\Delta(\rho\gamma)$ 
defined
above should be zero. 
Our numerical result for isospin effects is
\begin{eqnarray}
\Delta(\rho\gamma)=-(5.4\pm 5.4)\% .
\label{107}
\end{eqnarray}

\section{Conclusion}
\label{conclusion}
In this paper, we calculated the branching ratio, direct CP
asymmetry and isospin breaking effects within the standard model
using the pQCD approach.
Our predictions for the physical quantities are summarized in
Tab.\ref{con}.

\begin{table}
\begin{center}
Numerical results\vspace{3mm}\\
\begin{tabular}{c}\hline\hline
Branching ratio\\
 $\begin{array}{ccccc}
 B^0\to\rho^0\gamma && B^0\to\omega\gamma &&B^+\to\rho^+\gamma\\
 (1.2\pm 0.7)\times 10^{-6}&& (1.1\pm 0.6)\times 10^{-6}&&(2.5\pm
  1.5)\times 10^{-6}
 \end{array}$\\
 \hline
 Direct CP asymmetry\\
 $\begin{array}{ccccc}
 B^0\to \rho^0\gamma && B^0\to\omega\gamma &&
B^+\to\rho^+\gamma\\
(17.6\pm 15.0)\%~~~~ &&(17.9\pm 15.2)\% ~~~~&& (17.7\pm 15.0)\%~~~~
 \end{array}$ \\
 \hline
  Isospin breaking effects\vspace{1mm}\\
 $\begin{array}{c}
\Delta (\rho\gamma)=-(5.4\pm 5.4)\%
\end{array}$\\
\hline
\end{tabular}
\caption{The conclusion related to the branching ratio, 
CP asymmetry, and isospin breaking effects. }
\label{con}
\end{center}
\end{table}

The 
${B\to \rho}$
magnetic form factor ${T_1^{\rho}}$ 
is defined as
\begin{eqnarray}
&&\hspace{-1cm}\langle \rho(P_2,\epsilon_{K^*})\mid iq^{\nu}\bar d  \sigma_{\mu\nu}b 
\mid B(P_1)\rangle 
=-iT_1^{\rho}(0)\epsilon_{\mu\alpha\beta\rho}\epsilon_{\rho}^{\alpha}P^{\beta}
q^{\rho}
\end{eqnarray}
where ${P=P_1+P_2}$, ${q=P_1-P_2}$, 
and the value computed by the pQCD approach is
${T_1^{\rho}=}$
${0.26\pm 0.07}$.
The value from the light-cone-QCD sum rule (LCSR) is
${T_1^{\rho}=0.29\pm 0.04}$ \cite{Ball:1998kk},
then our value of the form factor is
in good agreement with LCSR.
The branching ratios only from  ${O_{7\gamma}}$
become
${Br(B^{\pm}\to\rho^{\pm}\gamma)=(2.4\pm 1.2)\times 10^{-6}}$,
${Br(B^{0}\to\rho^{0}\gamma)=(1.1\pm 0.5)\times 10^{-6}}$,
and\\
 ${Br(B^{0}\to\omega\gamma)=(1.0\pm 0.5)\times 10^{-6}}$;
then by comparing them to Eqs.(\ref{eq2})-(\ref{eq1}),
we can see that ${O_{7\gamma}}$ contributions
are dominant.

The subtle excess of the
branching ratio of ${B^0\to\rho^0\gamma}$
compared to that of ${B^0\to\omega\gamma}$ 
is caused by the following two reasons:
(1)
the difference in the meson mass
and decay constants between ${\rho}$
and ${\omega}$;
(2)
the annihilation contributions
from ${O_1}$ to ${O_6}$.
We examined these possibilities,
and concluded that the subtle excess of the
branching  ratio
mainly comes from (1),
and 
the  effects from (2) are very small.

The isospin breaking effect ${\Delta(\rho\gamma)}$
is caused by the contributions
${O_{8g}}$ (Fig.\ref{figo8g}), charm and up quark
loop contributions (Fig.\ref{loopa}), and
${O_1\sim O_6}$ annihilation contributions
(Figs.\ref{anni1}-\ref{anni3}).
Our prediction for this quantity is  given 
in Eq.(\ref{107}).
The main contributions to the isospin breaking effect come
from ${O_1\sim O_6}$ annihilation diagrams.
In general, we can expect that
 the annihilation contributions
are suppressed by the factor  ${O(m_q/m_b)}$, 
where ${m_q=m_u,~m_d}$.
In our computation, 
weak annihilations caused by ${O_3\sim O_6}$
are about 5\% and tree annihilations caused by ${O_1, O_2}$
are 1\% in the neutral modes, (see Tabs.\ref{rho0} and \ref{omega});
on the other hand, in the charged mode, weak annihilations are about 2\%
and tree annihilations are about 20\%; this contribution
is large because it is a color allowed process (see Tab.\ref{chargedrho})
in the amplitudes. 
If we neglect ${O_1\sim O_6}$ annihilation contributions,
the isospin breaking effects
have the opposite sign:
${\Delta(\rho\gamma)=+(3.4\pm 3.4)\%}$.
Thus
the
annihilation contributions are crucial
to the isospin breaking effects.

When we compare our results with the world averages of 
experimental data for the
${b\to d\gamma}$ decay modes
\cite{Group(HFAG):2005rb},
our results for the branching ratios are somewhat large.
For now, we shall not worry about it for the following reason:
Note that
our conclusion given in Tab.\ref{con},
\begin{eqnarray}
Br(B^0\to \rho^0\gamma)\approx Br(B^0\to \omega\gamma)\approx
\frac{1}{2}Br(B^+\to \rho^+\gamma),
\label{I}
\end{eqnarray}
follows from the isospin symmetry and the fact that contribution 
from the ${O_{7\gamma}}$
operator dominates over all other contributions.
A similar conclusion has been derived from the ${B\to K^*\gamma}$
decay mode \cite{Keum:2000wi}, and
experimental results for ${B\to K^*\gamma}$
agree with our conclusions.
While the error is large,
the relationships indicated by Eq.(\ref{I})
are not obviously seen in the recent experimental data.
We thus feel it is too early to discuss the validity of 
Eq.(\ref{I}).
We expect that the data may
change by about a factor two
if Eq.(\ref{I})
is approximately valid.

\section*{Acknowledgments}
The work of C.D.L\"u and M.Z.Yang is supported in part by
National   Science Foundation of China. 
A.I.S acknowledges support from JSPS Grant No.C-17540248.
We thank H.N.Li for
helpful communication and discussions.

\appendix
\section{Wave function}
\label{C}
In the calculation of the decay amplitude, the wave functions of
meson states can be defined via nonlocal matrix elements of quark
operators sandwiched between meson states and vacuum. Next let us
introduce the wave functions needed in this work.

The two leading-twist B meson wave functions can be defined
through the following nonlocal matrix element \cite{Bwave}:
\begin{eqnarray}
&&\int_0^1\frac{d^4z}{(2\pi)^4}e^{i\bf{k_1}\cdot z}
   \langle 0|\bar{q}_\alpha(z)b_\beta(0)|\bar B(p_B)\rangle
   \nonumber\\
&&\hspace{0.5cm}=\frac{i}{\sqrt{2N_c}}\left\{(\not p_B+M_B)\gamma_5
\left[\frac{\not v }{\sqrt{2}}\phi_B^+ ({\bf k_1})+\frac{ \not
n}{\sqrt{2}} {\phi}_B^-({\bf k_1})\right]\right\}_{\beta\alpha}
   \nonumber\\
&&\hspace{0.5cm}=-\frac{i}{\sqrt{2N_c}}\left\{(\not p_B+M_B)\gamma_5
\left[\phi_{B} ({\bf k_1})+ \sqrt{2}{\not n} \bar{\phi}_{B}({\bf
k_1})\right]\right\}_{\beta\alpha}, \nonumber\\
\label{aa1}
\end{eqnarray}
where $\bf{k_1}$ is the momentum of the light quark in the $B$ meson,
and $n= (1,0,{\bf 0_T})$, and $v=(0,1,{\bf 0_T})$. The
normalization conditions for these two wave functions are
\begin{equation}
\int d^4 k_1\phi_B^+({\bf k_1})=\frac{f_B}{2\sqrt{2N_c}}, ~~~ \int
d^4 k_1 \phi_B^-({\bf k_1})=\frac{f_B}{2\sqrt{2N_c}}.
\end{equation}
The relations between $\phi_{B}$, $\bar{\phi}_{B}$ and $\phi_B^+$,
$\phi_B^-$ are
\begin{equation}
  \phi_{B} =\phi_B^+~,~~  \bar{\phi}_{B}   = \frac{\phi_B^+ -\phi_B^-}{2}  .
  \label{phiB}
\end{equation}

In practice it is convenient to work in the impact parameter $b$
space rather than the transverse momentum space
($k_{\bot}$-space). So we make a Fourier transformation $\int
d^2k_\bot e^{-i{\bf k}_\bot \cdot {\bf b}}$ to transform the wave
functions and hard amplitude into $b$-space. Then the
normalization condition for $\phi_{B}$ and $\bar{\phi}_{B}$ can be
expressed as
\begin{equation}
\int_0^1 dx \phi_B(x,b=0)=\frac{f_B}{2\sqrt{2N_c}}, ~~~ \int_0^1
dx \bar{\phi}_B(x,b=0)=0\; .
\end{equation}
$\phi_{B}$ and $\bar{\phi}_{B}$ include bound state effects; they
are controlled by nonperturbative dynamics. They can be treated
by models or by solving the equation of motion in heavy quark
limit. 

 In some particular models,
$\phi_B$ and $\bar{\phi}_{B}$ can be selected such that the
contribution of $\bar{\phi}_{B}$ is the next-to-leading-power
$\bar{\Lambda}/m_B$ \cite{kls}. In this case the contribution of
$\bar{\phi}_{B}$ can be neglected at leading-power. Hence, only
$\phi_{B}$ is considered in this case. We adopt the model for
$\phi_{B}$ in the impact parameter $b$ space, which is widely used
in the study of $B$ decays in the perturbative QCD approach
\cite{pQCD}
\begin{equation}
\phi_{B}(x,b)=N_B
x^2(1-x)^2\mathrm{exp}\left[-\frac{1}{2}\left(\frac{xM_B}{\omega_B}\right)^2
-\frac{\omega_B^2b^2}{2}\right]\; ,
\end{equation}
where the shape parameter $\omega_B$ has been determined as
$\omega_B=0.40\mathrm{GeV}$, and $N_B$ is the normalization
constant.

In $B\to\rho\gamma$ and $\omega\gamma$ decays, $\rho$ and $\omega$
meson can only be transversely polarized. We only need to consider
the wave function of transversely polarized $\rho$ or $\omega$
meson. They are defined by
\begin{eqnarray}
&&\hspace{-8mm}
\langle\rho/\omega(P,\epsilon_V^T)|\bar{d_{\alpha}}(z)u_{\beta}(0)|0\rangle
=\frac{1}{\sqrt{2N_c}}\int_0^1 dx e^{ixP\cdot z}\Big[
M_{V}[\Slash{\epsilon}_{V}^{\ast T}]\phi_{V}^v(x)\nonumber\\
&&\hspace{4.5cm}
+[\Slash{\epsilon}_{V}^{\ast T}\hspace{1mm}\Slash{P}]\phi_{V}^T(x)
-\frac{M_{V}}{P\cdot n_+}i\epsilon_{\mu\nu\rho\sigma}
 [\gamma^5\gamma^{\mu}]\epsilon_{V}^{T
 \nu}P^{\rho}n_+^{\sigma}\phi_{V}^a(x)
\Big].\hspace{10mm}
\end{eqnarray}
For the transverse $\rho$ meson, the distribution amplitudes are
given as \cite{ball}:
    \begin{eqnarray}
\phi_{\rho}^T(x) &=& \frac{3f_{\rho}^T }{\sqrt{6} }
  x (1-x)  \left[1+ 0.2C_2^{3/2} (t) \right],
 \\
    \phi_{\rho}^v(x) &=&  \frac{f_{\rho} }{2\sqrt{6} }
  \Big[  \frac{3}{4} ( 1+ t^2) +0.24( 3 t^2-1) \nonumber\\
&&\hspace{1.5cm}+0.12 ( 3-30 t^2
  +35 t^4) \Big],\\
\phi_{\rho}^a(x) &=&  \frac{3 f_{\rho}}{4\sqrt{6}
 }~  t  \left[1+ 0.93 (10 x^2 -10 x +1) \right] ,
\end{eqnarray}
where $t=1-2x$
and 
\begin{eqnarray}
\phi_{V}^T(x)=\frac{f_{V}^T}{2\sqrt{2N_c}}\phi_{\perp},
\hspace{8mm}\phi_{V}^v(x)=\frac{f_{V}}{2\sqrt{2N_c}}g_{\perp}^{(v)},
\hspace{8mm}
\phi_{V}^a(x)=\frac{f_{V}}{8\sqrt{2N_c}}
\frac{d}{dx}g_{\perp}^{(a)}.\nonumber
\end{eqnarray}
We use ${\epsilon_{0123}=1}$ and
set the normalization condition
about ${ \phi_i =\{\phi_{\perp},g_{\perp}^{(v)},g_{\perp}^{(a)}\}}$ 
as
\begin{eqnarray}
\int_0^1dx \phi_i(x)=1.
\end{eqnarray}
The Gegenbauer polynomial is defined by
 \begin{equation}
 C_2^{3/2} (t) = \frac{3}{2} (5t^2-1).
 \end{equation}

\section{Some Functions}
\label{B}
The expressions for some functions are presented in this appendix.
In our numerical calculation, we
use the leading order ${\alpha_s}$ formula as
\begin{eqnarray}
\hspace{-5mm}\alpha_s(\mu)=\frac{2\pi}{\beta_0\ln(\mu/\Lambda_{n_f})},\hspace{5mm}
\beta_0 &=&\frac{33-2n_f}{3},
\end{eqnarray}
and we fix the number of the flavor as ${n_f=4}$.
The explicit expression for 
Sudakov factor ${s(t,b)}$ is given by \cite{Botts:kf} as follows:
\begin{eqnarray}
s(t,b)&=&\int^t_{1/b}\frac{d\mu}{\mu}
\left[\ln{\left(\frac{t}{\mu}\right)}
A(\alpha_s(\mu))+B(\alpha_s(\mu))
\right],\\
A&=&C_F \frac{\alpha_s}{\pi}+
\left(\frac{\alpha_s}{\pi}\right)^2
\left[\frac{67}{9}-\frac{{\pi}^2}{3}-\frac{10}{27}n_f +
\frac{2}{3}\beta_0\ln \left(\frac{e^{\gamma}_E}{2}\right)
\right],\\
B&=&\frac{2}{3}\frac{\alpha_s}{\pi}\ln\left(\frac{e^{2\gamma_E}-1}{2}\right),\\
\nonumber
\end{eqnarray}
where ${\gamma_E =0.5722}$ is Euler constant and ${C_F=4/3}$ is color factor.
The meson wave function including summation factor
has energy dependence,
\begin{eqnarray}
\phi_B(x_1,b_1,t)&=&\phi_B(x_1,b_1)\exp{[-S_B(t)]},\\
\phi_{V}(x_2,t)&=&\phi_{V}(x_2)\exp{[-S_{V}(t)]},
\end{eqnarray}
and the total functions including Sudakov factor
and ultraviolet divergences are
\begin{eqnarray}
S_B(t)&=&s(x_1P_1^-,b_1)+2\int^t_{1/b_1}\frac{d\bar{\mu}}{\bar{\mu}}\gamma
(\alpha_s(\bar{\mu})),\\
S_{V}(t)&=&s(x_2P_2^-,b_2)+s((1-x_2)P_2^-,b_2)
+2\int^t_{1/b_2}
\frac{d\bar{\mu}}{\bar{\mu}}\gamma(\alpha_s(\bar{\mu})).
\end{eqnarray}
Threshold factor is expressed as
below
\cite{Keum:2000wi, Li:2001ay},
and  we take the value ${c=0.4}$:
\begin{eqnarray}
S_t(x)=\frac{2^{1+2c}\Gamma(3/2+c)}{\sqrt{\pi}\Gamma(1+c)}
[x(1-x)]^c.
\end{eqnarray}

\end{document}